\documentclass[twocolumn]{aastex631}

\usepackage[utf8]{inputenc}
\usepackage{hyperref}
\usepackage{acronym}   
\usepackage{xspace}
\usepackage{amsmath}
\usepackage{amssymb}
\usepackage{xcolor}
\usepackage{physics}
\usepackage{chngcntr}
\usepackage[figuresright]{rotating}
\usepackage[caption=false]{subfig}
\usepackage{graphicx}
\hypersetup{pdfauthor={Name}}

\newcommand{\Msun}{\ensuremath{\xspace\rm{M}_{\odot}}\xspace}

\newcommand{\qCrit}{\ensuremath{q_c}\xspace}
\newcommand{\qCritGeTwenty}{\ensuremath{q_{c,\;\mathrm{Ge20}}}\xspace}
\newcommand{\qCritClaeys}{\ensuremath{q_{c,\;\textrm{Claeys14}}}\xspace}
\newcommand{\qCritGeTwentyIC}{\ensuremath{\tilde{q}_{c,\;\mathrm{Ge20}}}\xspace}
\newcommand{\zetaSPH}{\ensuremath{\zeta_\mathrm{Comp}}\xspace}
\newcommand{\zetaGeTwenty}{\ensuremath{\zeta_\mathrm{Ge20}}\xspace}
\newcommand{\zetaGeTwentyIC}
{\ensuremath{\tilde{\zeta}_\mathrm{Ge20}}\xspace}
\newcommand{\zetaClaeys}{\ensuremath{\zeta_\mathrm{Claeys14}}\xspace}
\newcommand{\betaCompas}{\ensuremath{\beta_{\mathrm{Comp}}}\xspace}
\newcommand{\fGamma}{\ensuremath{f_{\gamma}}\xspace}
\newcommand{\aPre}{\ensuremath{a_\mathrm{pre}}\xspace}
\newcommand{\aPost}{\ensuremath{a_\mathrm{post}}\xspace}
\newcommand{\aPostOnAPre}{\ensuremath{\aPost/\aPre}\xspace}
\newcommand{\fSMT}{\ensuremath{f_\mathrm{SMT}}\xspace}

\newcommand{\zetaRL}{\ensuremath{\zeta_{L}}\xspace}

\acrodef{AM}{angular momentum}
\acrodef{ZAMS}{zero-age main sequence}
\acrodef{MS}{main sequence}
\acrodef{HG}{Hertzsprung gap}
\acrodef{NS}{neutron star}
\acrodef{DNS}{double neutron star}
\acrodef{BH}{black hole}
\acrodef{BBH}{binary black hole}
\acrodef{WD}{white dwarf}
\acrodef{CO}{compact object}
\acrodef{MT}{mass transfer}
\acrodef{SMT}{stable mass transfer}
\acrodef{IMF}{initial mass function}
\acrodef{UMT}{unstable mass transfer}
\acrodef{GW}{gravitational wave}
\acrodef{RLOF}{Roche-lobe overflow}
\acrodef{CEE}{common envelope evolution}
\acrodef{BPS}{rapid binary population synthesis}
\acrodef{HMXB}{high-mass X-ray binary}
\acrodefplural{HMXB}{high-mass X-ray binaries}
\acrodef{SN}{supernova}
\acrodefplural{SN}[SNe]{supernovae}
\acrodef{CCSN}{core-collapse supernova}
\acrodefplural{CCSN}[CCSNe]{core-collapse supernovae}
\acrodef{SESN}{stripped-envelope supernova}
\acrodefplural{SESN}[SESNe]{stripped-envelope supernovae}
\acrodef{ECSN}{electron-capture supernova}
\acrodefplural{ECSN}[ECSNe]{electron capture supernovae}

\defcitealias{Ge_etal.2020_AdiabaticMassLoss}{Ge20}

\shortauthors{Willcox et al.}
\shorttitle{Stable MT}

\begin{document}

\title{The Impact of Angular Momentum Loss on the Outcomes of Binary Mass Transfer}
\date{August 2023}

\correspondingauthor{Reinhold Willcox}

\author[0000-0003-0674-9453]{Reinhold Willcox}
\affiliation{School of Physics and Astronomy
Monash University,
Clayton, VIC 3800, Australia}
\affiliation{The ARC Centre of Excellence for Gravitational Wave Discovery -- OzGrav, Australia}
\email{reinhold.willcox@monash.edu}

\author[0000-0002-1417-8024]{Morgan MacLeod}
\affiliation{Center for Astrophysics \textbar Harvard \textsl{\&}  Smithsonian 60 Garden Street, MS-16, Cambridge, MA 02138, USA}

\author[0000-0002-6134-8946]{Ilya Mandel}
\affiliation{School of Physics and Astronomy
Monash University,
Clayton, VIC 3800, Australia}
\affiliation{The ARC Centre of Excellence for Gravitational Wave Discovery -- OzGrav, Australia}

\author[0000-0002-8032-8174]{Ryosuke Hirai}
\affiliation{School of Physics and Astronomy
Monash University,
Clayton, VIC 3800, Australia}
\affiliation{The ARC Centre of Excellence for Gravitational Wave Discovery -- OzGrav, Australia}

\begin{abstract}

We use the rapid binary population synthesis code COMPAS to investigate commonly used prescriptions for the determination of mass transfer stability in close binaries and the orbital separations after stable mass transfer. The degree of orbital tightening during non-conservative mass transfer episodes is governed by the poorly-constrained angular momentum carried away by the ejected material.  Increased orbital tightening drives systems towards unstable mass transfer leading to a common envelope. We find that the fraction of interacting binaries that will undergo only stable mass transfer throughout their lives fluctuates between a few and $\sim 20\%$ due to uncertainty in the angular momentum loss alone. If mass transfer is significantly non-conservative, stability prescriptions that rely on the assumption of conservative mass transfer under-predict the number of systems which experience unstable mass transfer and stellar mergers. This may substantially impact predictions about the rates of various transients, including luminous red novae, stripped-envelope supernovae, X-ray binaries, and the progenitors of coalescing compact binaries.

\end{abstract}


\section{Introduction}
\label{sec:intro}

Stellar binaries that interact via \ac{MT} during their evolution produce a wide variety of astrophysically interesting objects and transients. Despite their central importance, these mass transfer phases are not yet fully understood. The majority of massive stars, progenitors of \acp{NS} and \acp{BH}, are born in close binaries that will interact at some point during their evolution \citep{Sana_etal.2012_BinaryInteractionDominates, Moe_DiStefano.2017_MindYourPs, DeMarco_Izzard.2017_DawesReviewImpact}. Typically, mass is transferred from an expanding donor star onto an accreting companion. The rate of \ac{MT} generally depends on the geometry of the system, along with the thermodynamic and hydrodynamic behavior of the transferred material. To date, studies can model either the 3D hydrodynamics or the thermodynamics of individual systems, but computational limitations preclude integration of the two effects and generalization to the entire parameter space of interacting binaries \citep{Ivanova_Nandez.2016_CommonEnvelopeEvents, MacLeod_Loeb.2020_PrecommonenvelopeMassLoss, MacLeod_Loeb.2020_RunawayCoalescencePrecommonenvelope, Marchant_etal.2021_RoleMassTransfer}.

In a somewhat simplified treatment, we often distinguish between stable mass transfer and unstable mass transfer. Although precise criteria for the onset of instability in mass transfer episodes have remained elusive, unstable mass transfer generally involves the engulfment of the accretor by the donor's envelope, resulting in either a rapid contraction of the orbit or a stellar merger. This is commonly known as the common envelope phase, or \ac{CEE}, and, in cases where the stars avoid merging, involves the loss of a large fraction of the donor’s envelope on a dynamical timescale \citep{Paczynski.1976_CommonEnvelopeBinaries}.

By contrast, stable \ac{MT} proceeds on a thermal or nuclear timescale and (assuming it does not immediately precede a phase of unstable \ac{MT}) always results in a detached binary. However, even this slower phase of mass transfer is sensitive to the donor's stellar structure -- in particular its density and entropy profiles -- which determine the donor's response to mass loss. The binary's trajectory through all later stages of binary evolution critically depends on this determination of stability and the binary separation following the interaction.

The boundary between stable and unstable \ac{MT} represents one of the major ongoing questions in binary evolution theory. Simplified, early treatments for the stability boundary are based on the response of the donor radius at the onset of \ac{MT} and account for the effects of non-conservative \ac{MT} \citep{Hjellming_Webbink.1987_ThresholdsRapidMass, Soberman_etal.1997_StabilityCriteriaMass}. In the case of non-conservative \ac{MT}, mass is removed from the binary along with some of the orbital angular momentum, though accurately computing the amount of removed angular momentum is challenging \citep{MacLeod_etal.2018_BoundOutflowsUnbound,MacLeod_etal.2018_RunawayCoalescenceOnset} and sensitive to the accretion mechanism \citep{Kalogera_Webbink.1996_FormationLowMassXRay, Goodwin_Woods.2020_BinaryEvolutionSAX}. More modern stability criteria track the entire donor structural response to mass loss on different timescales, but in order to keep the parameter space manageable, these typically assume fully conservative \ac{MT} \citep{Ge_etal.2020_AdiabaticMassLoss, Woods_Ivanova.2011_CanWeTrust, Pavlovskii_etal.2017_StabilityMassTransfer,  Temmink_etal.2022_CopingLossStability}.

In this study, we explore the consequences of implementing these newer stability criteria at the expense of flexibility in the \ac{MT} accretion efficiency. We use \ac{BPS} models to simulate the evolution of stellar binaries using a variety of prescriptions for \ac{MT} stability, focusing particularly on the \ac{MT} efficiency and angular momentum loss. 

The paper is structured as follows. In Sec.~\ref{sec:methodology}, we review the analytics of mass exchange in binary stars and the distinction between stable and unstable \ac{MT}, and introduce our population synthesis models. In Sec.~\ref{sec:results}, we present the results of our simulations: the fraction of each population which experiences only stable \ac{MT} at different evolutionary epochs, and the separation distributions following stable \ac{MT}. In Sec.~\ref{sec:discussion_conclusion}, we discuss which parameter variations are the most impactful and implications for future studies.

\section{Methodology}
\label{sec:methodology}

\subsection{Analytics of stable mass transfer}
\label{sec:smt_analytics}

We review the analytics for binary \ac{MT} and define the relevant free parameters. We borrow heavily from the didactic approach of \citet{Pols.2018_CourseNotesBinary} as well as \citet{Soberman_etal.1997_StabilityCriteriaMass}. For the simplified case of a circular, co-rotating binary, the stellar material around each star is gravitationally bound to its host if the stellar radius does not exceed its Roche lobe, the equipotential surface passing through the L1 point. If the star expands beyond its Roche lobe, gas will flow through a nozzle near the L1 point and fall toward the companion star. The volume-equivalent Roche-lobe radius $R_L$ of the donor is  

\begin{equation}\label{eq:roche_lobe}
R_L =  \frac{0.49 a}{0.6 + q^{2/3} \ln(1 + q^{-1/3})}, 
\end{equation}
where $a$ is the orbital separation, and $q=M_a/M_d$ is the binary mass ratio in terms of the mass of the accretor $M_a$ and the donor $M_d$ \citep{Eggleton.1983_AproximationsRadiiRoche}. If the donor radius $R_d > R_L$, mass transfer ensues via \ac{RLOF}. 

For a binary in a circular orbit, the orbital \ac{AM} is 
\begin{equation}\label{eq:angmom}
J = M_a M_d \sqrt{\frac{Ga}{M_a+M_d}},
\end{equation}
where G is the gravitational constant. If the orbit is eccentric, \ac{MT} may begin in bursts near the point of closest approach; we assume that \ac{MT} drives the binary to circularize at periapsis (though see \citet{Sepinsky_etal.2007_InteractingBinariesEccentric, Sepinsky_etal.2009_InteractingBinariesEccentric, Sepinsky_etal.2010_InteractingBinariesEccentric, Dosopoulou_Kalogera.2018_OrbitalEvolutionMassTransferring}).

As \ac{MT} proceeds, some fraction $\beta$ of the mass lost from the donor will be accreted by the companion, 

\begin{equation}
\dot{M_a} = -\beta \dot{M_d}.
\end{equation}
The MT efficiency $\beta$ in general depends on the properties of the binary prior to the \ac{MT} and changes with time, though it is often assumed to be a constant for a given \ac{MT} episode. $\beta=1$ corresponds to conservative \ac{MT}, in which all material lost from the donor is accreted, while $\beta < 1$ refers to non-conservative \ac{MT}, and $\beta=0$ is fully non-conservative. 

Any mass that is not accreted is ejected from the binary system, extracting orbital angular momentum. Traditionally, this is represented by the specific \ac{AM} $\gamma$ of the ejected material in units of the orbital specific \ac{AM}. For simplicity, we introduce a new parametrization for the \ac{AM} loss, the effective decoupling radius $a_\gamma$.  This can be thought of as the distance from the center of mass at which matter is ejected. The parameters $a_\gamma$ and $\gamma$ are related by

\begin{equation}\label{eq:gamma_definition}
\begin{split}
\gamma
& = \left(\frac{\dot{J}}{\dot{M_a}+\dot{M_d}} \right)  \bigg/
\left(\frac{J}{M_a+M_d} \right)   \\
& = \left(a_\gamma^2 \omega_{orb} \right) \bigg/
\left(\frac{J}{M_a+M_d} \right)   \\
& = \left(\frac{a_\gamma}{a}\right)^2 \frac{(1 + q)^2}{q}, \\
\end{split}
\end{equation}
where $\omega_{orb}$ is the orbital angular frequency. 

As with the efficiency parameter $\beta$, $\gamma$ is generally dependent on time and attributes of the binary, such as the mass ratio and degree of rotational synchronization \citep{MacLeod_Loeb.2020_PrecommonenvelopeMassLoss, MacLeod_Loeb.2020_RunawayCoalescencePrecommonenvelope}, but is similarly often simplified to a constant. For matter which is removed from the vicinity of the donor star (the ``Jeans'' mode for a fast, isotropic wind), $a_\gamma$=$aq/(1+q)$, so $\gamma=q$. Material which instead decouples from the binary near the accretor has $a_\gamma=a/(1+q)$ and $\gamma = 1/q$ (the ``isotropic re-emission'' mode). The maximal specific \ac{AM} loss for co-rotating matter occurs at the L2 point, $ a_\gamma \approx a_{\text{L2}} \lesssim 1.25a$ (see Fig.~\ref{fig:zetaRL_heatmap})\footnote{Lagrange points were calculated using the methodology described in \href{https://map.gsfc.nasa.gov/ContentMedia/lagrange.pdf}{https://map.gsfc.nasa.gov/ContentMedia/lagrange.pdf}}. Larger specific \ac{AM} loss may be achieved if the ejected matter forms a circumbinary ring which applies a torque on the orbit, though we do not consider this case here.  

Taking the time derivative of Eq.~(\ref{eq:angmom}) and re-arranging with the definitions of $\beta$ and $\gamma$, the relative change in separation is

\begin{equation}\label{eq:separation_change}
    \frac{\dot{a}}{a} = 
    -2 \frac{\dot{M_d}}{M_d} 
    \left[ 
    1 - \frac{\beta}{q} + (\beta-1)\left(\gamma+\frac{1}{2}\right)
    \frac{1}{1+q} 
    \right].
\end{equation}

\subsubsection{Stability criteria}
\label{sec:stability_criteria}

To predict the outcome of the binary post-interaction, we distinguish mass transfer which leads to \ac{CEE} from that which does not. Using Eqs.~(\ref{eq:roche_lobe}) and (\ref{eq:separation_change}), we define the response of the donor's Roche lobe to mass loss by the logarithmic derivative 

\begin{sidewaysfigure*}
\centering
\vspace{10cm}
\includegraphics[width=\columnwidth]{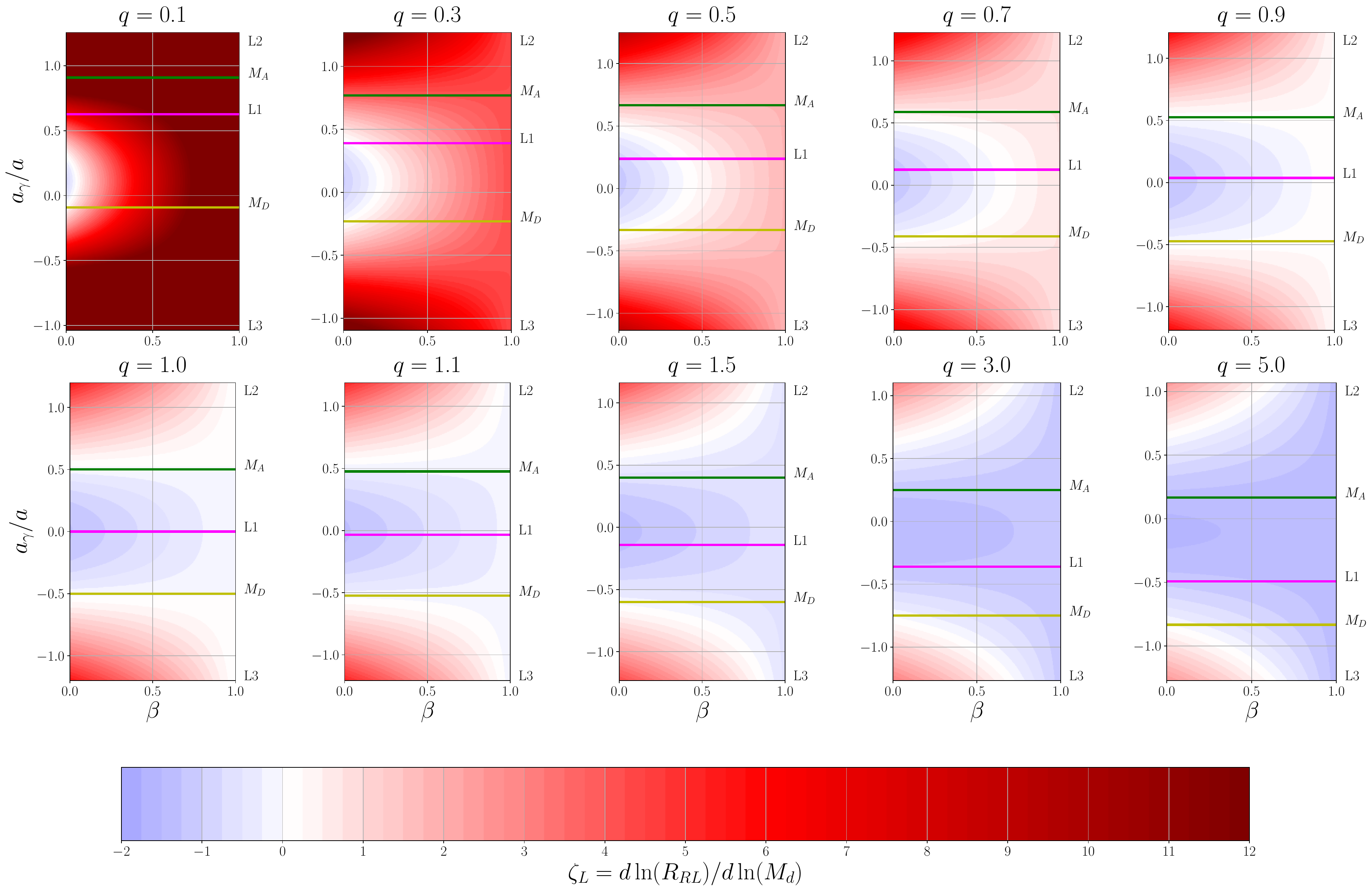}
\caption{
\textbf{The Roche-lobe response to mass transfer $\zetaRL$} for different values of the mass ratio $q=M_a/M_d$, efficiency $\beta$, and specific angular momentum of non-accreted material. Here, the specific \ac{AM} is parametrized by $a_\gamma$, the distance from the binary center-of-mass at which the non-accreted material decouples from the system, in units of the semi-major axis $a$. The light and dark green horizontal lines mark the positions of the donor and accretor, respectively, with the donor always positioned toward the bottom. The upper and lower bounds correspond to the positions of the L2 and L3 Lagrangian points, respectively, which slowly vary in absolute value between $\sim$[1.2, 1.275]~$a$ for this range in $q$. The pink horizontal line represents the L1 point, and $a_\gamma/a = 0.0$ demarcates the binary center-of-mass. The range considered for \ac{AM} loss in our proxy parameter \fGamma, thus represents the space between the dark green $M_A$ line and the top of the axis. 
}
\label{fig:zetaRL_heatmap}
\end{sidewaysfigure*}

\begin{equation}\label{eq:zeta_rl}
\begin{split}%[r|c|l]
    \zetaRL  = & \pdv{\ln R_{L}}{\ln M_d}, \\
     = & \pdv{\ln a}{\ln M_d} + \pdv{\ln (R_L/a)}{\ln q} \pdv{\ln q}{\ln M_d} \\
     = & -2 \left[ 1 - \frac{\beta}{q} + (\beta-1)\left(\gamma+\frac{1}{2}\right) \frac{1}{1+q} \right] + \\
     & \frac{1}{3} \left( 2 - 
     \frac{1.2 + (q^{-1/3} + q^{-2/3})^{-1}}
     {0.6 + q^{2/3}\ln(1+q^{-1/3})}  \right)
     \left(\frac{\beta}{q}+1\right).
\end{split}
\end{equation}
This is purely a function of $q$, $\beta$, and $\gamma$, and is explicitly independent of the mass transfer timescale. For most of the parameter space, mass transfer from a high mass donor to a low mass accretor will tighten the binary and shrink the Roche lobe, and vice versa. Eq.~(\ref{eq:zeta_rl}) is shown graphically in Fig.~\ref{fig:zetaRL_heatmap} as a function of these parameters, with $\gamma$ replaced by the decoupling separation $a_\gamma/a$ according to Eq.~(\ref{eq:gamma_definition}).

By contrast, the radial response of the donor to mass loss,

\begin{equation}\label{eq:zeta_star}
    \zeta_{*} = \pdv{\ln R_d}{\ln M_d},
\end{equation}
strongly depends on the mass transfer timescale, as well as the evolutionary phase of the donor. Following a small amount of mass loss, the stellar structure will react on the short dynamical timescale to restore hydrostatic equilibrium. This initial, adiabatic radial response is parametrized as $\zeta_{*} = \zeta_{ad}$.

Following \citet{Soberman_etal.1997_StabilityCriteriaMass}, we label mass transfer as dynamically unstable if the adiabatic response of the donor to mass loss at the beginning of mass transfer is to expand relative to its Roche lobe, $\zeta_* < \zeta_L$, leading to an increase in the mass loss rate and a runaway process on the dynamical timescale. According to this definition, unstable \ac{MT} leading to \ac{CEE} sets in only if the \ac{MT} occurs on the dynamical timescale at the onset of the mass exchange. Otherwise, the \ac{MT} is stable.

Eq.~(\ref{eq:zeta_star}) can be solved analytically for some simplified polytropic models \citep{Hjellming_Webbink.1987_ThresholdsRapidMass, Soberman_etal.1997_StabilityCriteriaMass}, or numerically using detailed 1D models of stars undergoing mass loss \citep{deMink_etal.2007_EfficiencyMassTransfer, Ge_etal.2020_AdiabaticMassLoss, Klencki_etal.2021_ItHasBe}. Broadly, the adiabatic response of the star is sensitive to the specific entropy gradient at the surface. For stars with radiative envelopes (intermediate and high mass stars on the \ac{MS} and early on the \ac{HG}), the specific entropy rises steeply near the surface.  Mass loss exposes lower entropy layers which contract rapidly ($\zeta_{ad} > 0$). As the mass loss continues, in the adiabatic regime (i.e. fixing the entropy profile) the entropy gradient tapers off and the donor contraction is moderated.

In stars with convective envelopes (low mass \ac{MS} stars and cool giants), energy is efficiently transported in convective layers leading to a flat specific entropy profile within the convective region. Unless the core mass constitutes $\gtrsim20\%$ of the total stellar mass, the adiabatic response of the donor to mass loss is to expand ($\zeta_{ad} < 0$) \citep{Soberman_etal.1997_StabilityCriteriaMass}. In the extreme case of a fully convective star, the response of the donor follows $R\propto M^{-1/3}$, i.e. $\zeta_{ad}=-1/3$ \citep{Hjellming_Webbink.1987_ThresholdsRapidMass, Kippenhahn_etal.2012_StellarStructureEvolution}.

However, \citet{Ge_etal.2010_AdiabaticMassLoss} argued that comparing $\zeta_*$ and $\zetaRL$ only at the onset of \ac{MT} may be a poor predictor of \ac{MT} stability, since both $\zeta_*$ and $\zetaRL$ change throughout the \ac{MT}. They found that while fully convective stars do expand rapidly during adiabatic mass loss, the outermost layers which extend beyond the Roche lobe are very diffuse, so that the actual mass loss rate is low. Although this is initially a runaway process, the mass loss rate may never exceed a (model dependent) critical value, in which case the \ac{MT} will proceed stably. 

Conversely, for some radiative donors, $\zeta_{ad}$ may be initially very high (rapid contraction), but decrease over time until $\zeta_{ad} \sim \zetaRL$, at which point the binary will experience what is known as the \emph{delayed dynamical instability} \citep{Hjellming_Webbink.1987_ThresholdsRapidMass, Ge_etal.2010_AdiabaticMassLoss}. As with the convective case, this delayed effect cannot be effectively captured from the initial stellar response at the onset of \ac{MT}, as defined in \citet{Soberman_etal.1997_StabilityCriteriaMass}, but requires a detailed treatment of the stellar response over the \ac{MT} episode.

\subsubsection{Critical mass ratios}
\label{sec:critical_mass_ratios}

Computing the evolution of $\zeta_*$ and $\zetaRL$ from detailed simulations for each \ac{MT} event is impractical for \ac{BPS} purposes. Instead, one can define a critical mass ratio \qCrit as the threshold mass ratio, prior to \ac{RLOF}, at which dynamical instability sets in \emph{at any point} during the mass transfer.  In practice, critical mass ratios are calculated from detailed models and depend on both the donor mass and radius (or, equivalently, evolutionary phase). Throughout this paper, we use the convention $q = M_a/M_d$, so that the mass transfer is unstable if $q<\qCrit$.

Because the \citet{Ge_etal.2010_AdiabaticMassLoss} definition includes these delayed effects, most critical mass ratio prescriptions tend to predict that mass transfer from radiative stars is more unstable, and from convective stars more stable, than is found in prescriptions that consider only the stellar response at the onset of \ac{MT}. However, some studies define critical mass ratios based on the initial \ac{MT} response and thus do not account for the delayed effects (e.g \citealt{Hurley_etal.2002_EvolutionBinaryStars, deMink_etal.2007_EfficiencyMassTransfer, Claeys_etal.2014_TheoreticalUncertaintiesType}).

Critical mass ratio prescriptions must implicitly specify a Roche-lobe response $\zetaRL$, i.e. a choice for the efficiency $\beta$ and \ac{AM} loss $\gamma$. For simplicity, this is usually taken to be $\beta=1$ (i.e. fully conservative mass transfer).  This, however, limits the applicability of such prescriptions, as we discuss later. Recent studies have invoked alternative definitions for stability, including overflow from the L2 point \citep{Pavlovskii_Ivanova.2015_MassTransferGiant, Pavlovskii_etal.2017_StabilityMassTransfer, Lu_etal.2022_RapidBinaryMass}, or rapid changes to the orbit \citep{Temmink_etal.2022_CopingLossStability}. However, these definitions also assume fully conservative \ac{MT}, and thus are similarly limited.

\subsection{Population synthesis}
\label{sec:popsynth}

We use the COMPAS\footnote{\url{http://github.com/TeamCOMPAS/COMPAS}} rapid binary population synthesis suite, including the fiducial parameter choices listed in \citet{TeamCOMPAS_etal.2022_COMPASRapidBinary}, with a few exceptions and recently added prescriptions detailed below. For each distinct model, we evolve $10^5$ binaries from \ac{ZAMS}. By default, we follow the traditional initial sampling distributions used in many \ac{BPS} codes. For the more massive primary, we draw its mass $M_1$ from the Kroupa \ac{IMF} $p(M_1) \propto M_1^{-2.3}$ between 5 and 100~\Msun \citep{Kroupa.2001_VariationInitialMass}. The mass ratio distribution is uniform $q \in [0.1, 1]$ (with no additional constraint imposed on the minimum of $M_2$), and the separation is drawn log-uniformly, $\log(a/\mathrm{AU}) \in [-2, 2]$. All single stars are included implicitly as very wide binaries for normalisation purposes. Since wide binaries do not interact, extending this upper limit is analogous to increasing the effectively single star fraction. Initial eccentricity is fixed at 0. We discuss variations to the initial parameter distributions in App.~\ref{sec:impact_correlated_ICs}.

Each binary is then evolved under the specified evolution model until one of the following termination conditions is reached: 

\begin{itemize}
    \item The binary experiences unstable mass transfer, 
    \item Either star experiences a \ac{SN}, or
    \item Both components evolve into white dwarfs.
\end{itemize}

We investigate the binaries at two epochs: immediately following the first mass transfer episode and following the final mass transfer episode prior to the termination condition, henceforth referred to as the End state. We classify binaries based on whether or not they have experienced only stable \ac{MT} up to the specified epoch. We additionally consider the orbital separations following the first \ac{MT} episode, if the \ac{MT} was stable. The outcome of \ac{CEE} is notoriously uncertain and the subject of ongoing research, so for simplicity we do not distinguish common-envelope survival from stellar mergers here \citep{Ivanova.2017_CommonEnvelopeProgress, Lau_etal.2022_CommonEnvelopesMassive, Hirai_Mandel.2022_TwostageFormalismCommonenvelope}.

By restricting our study to pre-\ac{SN} and pre-Double White Dwarf interactions, and by ignoring post-\ac{CEE} outcomes, we limit the scope of our parameter space and ensure that our conclusions are more robust. Many systems will not interact: this is largely dependent on the initial conditions (and, to a lesser extent, on the stellar winds at early phases). These non-interacting systems are generally ignored here, except where otherwise specified.

\subsection{Parameter variations}
\label{sec:parameter_variations}

\subsubsection{Roche-lobe response}
\label{sec:roche_response}

The rate of accretion onto a non-degenerate companion is often assumed to be restricted by the accretor's thermal timescale, $\dot{M_a} \sim M_a/\tau_{KH, a}$. We calculate the accretion efficiency as

\begin{equation}\label{eq:beta_ratio}
    \beta = -\frac{\dot{M_a}}{\dot{M_d}} = \min\left(\frac{C M_a/\tau_{KH, a}}{M_d/\tau_{KH, d}}, 1\right),
\end{equation}
where $\tau_{KH}$ is the Kelvin-Helmholtz timescale, and the pre-factor $C$ (10 in this study) is included to account for an increase in the maximum accretion rate due to expansion of the accretor \citep{Paczynski_Sienkiewicz.1972_EvolutionCloseBinaries, Neo_etal.1977_EffectRapidMass, Hurley_etal.2002_EvolutionBinaryStars, Schneider_etal.2015_EvolutionMassFunctions}. During Case A \ac{MT} -- defined when the donor is on the \ac{MS} -- $\beta \approx 1$ if the component masses are roughly comparable. During Case B/C \ac{MT} -- defined for more evolved donors -- $\beta$ is typically much closer to 0. Throughout, we follow the stellar type convention defined by \citet{Hurley_etal.2000_ComprehensiveAnalyticFormulae}.

We first consider the impact of uncertainties in the Roche-lobe response to mass loss, $\zetaRL$, by varying the \ac{MT} efficiency $\beta$ and the specific \ac{AM} of non-accreted matter $\gamma$. The default efficiency in COMPAS, \betaCompas, uses the ratio of thermal timescales defined in Eq.~(\ref{eq:beta_ratio}). We also include variations with fixed values of $\beta = \{0.0, 0.5, 1.0\}$. To investigate the effect of different \ac{AM} loss modes, we introduce a new parameter, $\fGamma \in [0, 1]$, which increases linearly with the decoupling radius between the accretor $a_{acc}$ and the L2 point $a_{L2}$,
\begin{equation}
    a_\gamma = a_{acc} + f_\gamma(a_{L2} - a_{acc}),
\end{equation}
such that $\fGamma=0$ corresponds to the isotropic re-emission model and $\fGamma=1$ corresponds to \ac{AM} loss from the L2 point. This range in \fGamma reflects results from hydrodynamic simulations, which indicate that non-accreted mass could decouple from the binary between the accretor and the L2 point, depending on the structural properties of the donor \citep{MacLeod_etal.2018_RunawayCoalescenceOnset, MacLeod_etal.2018_BoundOutflowsUnbound, MacLeod_Loeb.2020_RunawayCoalescencePrecommonenvelope}. Many rapid \ac{BPS} codes assume the isotropic re-emission model by default. 

Since the impact of $\gamma$ is moderated by $\beta$, we consider pairwise combinations of these parameters, noting that $\gamma$ is irrelevant for fully conservative mass transfer.

\subsubsection{Donor radial response}
\label{sec:donor_radial_response}

\begin{sidewaysfigure*}
\centering
\vspace{10cm}
\includegraphics[width=\columnwidth]{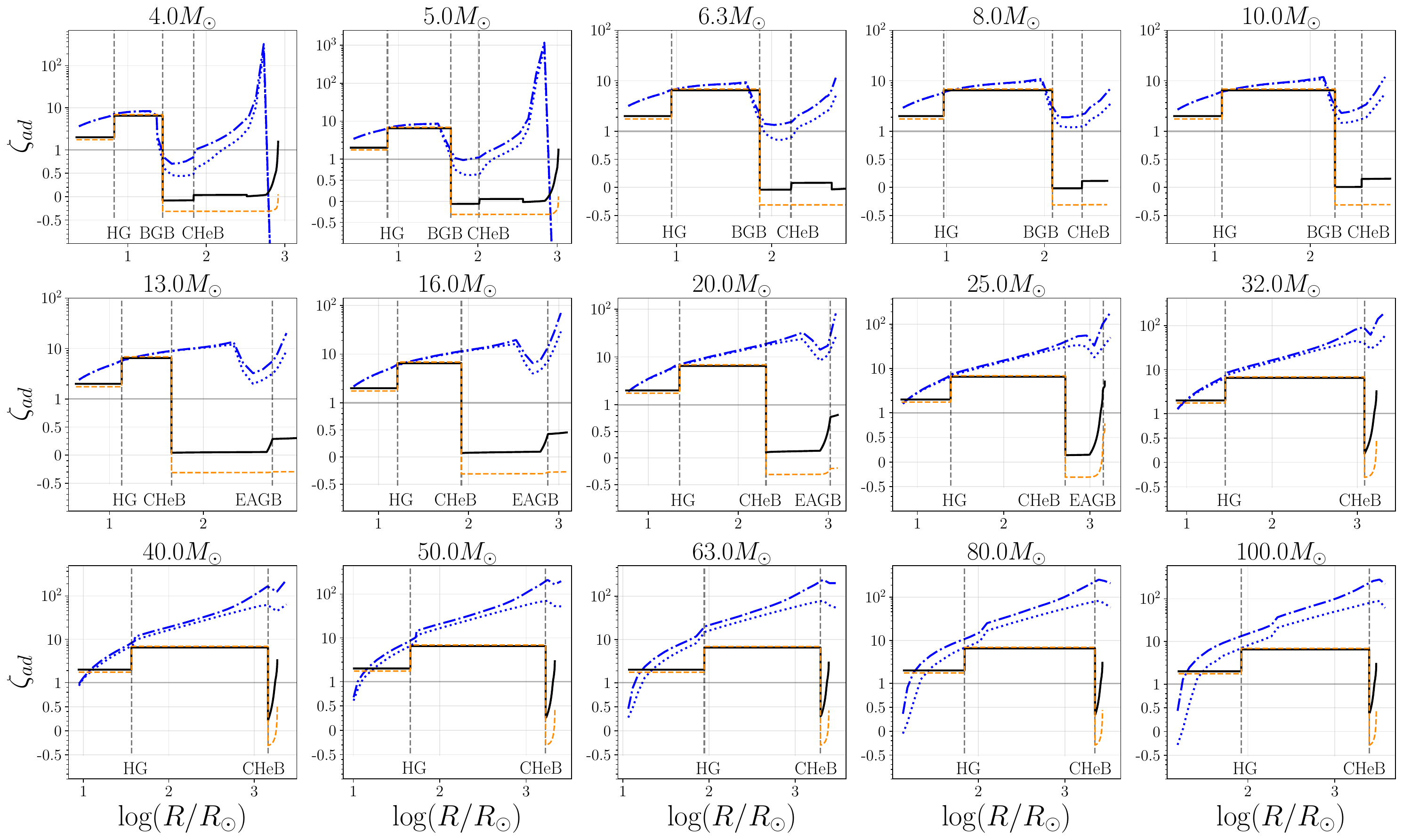}
\caption{
\textbf{The donor response to mass loss} $\zeta_{ad}$ as a function of donor radius, for selected \ac{ZAMS} masses.  The solid, black lines are the COMPAS default \zetaSPH (see text for details). The blue lines show \zetaGeTwenty (dotted, blue) and \zetaGeTwentyIC (dot-dashed, blue), from \citet{Ge_etal.2020_AdiabaticMassLoss}, corresponding to their models for fully adiabatic mass loss, and an artificially isentropic convective envelope, respectively. The dashed, orange lines represent the \zetaClaeys prescription \citep{Claeys_etal.2014_TheoreticalUncertaintiesType}. Values for $\zeta_{ad}$ here can be compared to those in Fig.~\ref{fig:zetaRL_heatmap} for $\zetaRL$ to determine MT stability. However, the $\zeta$ values from \citet{Claeys_etal.2014_TheoreticalUncertaintiesType} and \citet{Ge_etal.2020_AdiabaticMassLoss} are derived from their respective critical mass ratios, which implicitly require that the mass loss is fully conservative, and thus can only be compared to \zetaRL where $\beta=1$. Additionally, since \citet{Ge_etal.2020_AdiabaticMassLoss} account for the delayed dynamical instability, \zetaGeTwenty and \zetaGeTwentyIC do not represent the initial donor radial response to mass loss, but rather the \emph{effective} $\zeta_{ad}$ that should be compared against \zetaRL to determine stability. The vertical lines represent the radii in our models when the stellar type transitions to the named type. In practice, we interpolate in both the radii and masses. Note that the y-axis is linear below 1 and logarithmic above, to better distinguish the regions of interest.
}
\label{fig:zetaGrid}
\end{sidewaysfigure*}

We additionally consider variations to the radial response of the donor to mass loss $\zeta_{ad}$. In our default model both \ac{MS} and \ac{HG} stars (in the mass range of interest) have radiative envelopes for the entirety of the phase, with $\zeta_{ad,MS} = 2$ and $\zeta_{ad,HG} = 6.5$, respectively (these values are motivated by a constant approximation to the \citet{Ge_etal.2020_AdiabaticMassLoss} results, see \citet{TeamCOMPAS_etal.2022_COMPASRapidBinary}). Giants, by contrast, are assumed to have convective envelopes. By default, we follow the prescription outlined in  \citet{Soberman_etal.1997_StabilityCriteriaMass}. According to this prescription, mass loss leads to radial expansion unless the core mass fraction becomes non-negligible. We refer to this set of prescriptions encompassing all stellar types as \zetaSPH.  

We also include several prescriptions for stability based on critical mass ratios, from \citet{Claeys_etal.2014_TheoreticalUncertaintiesType} and \citet{Ge_etal.2020_AdiabaticMassLoss}. From \citet{Claeys_etal.2014_TheoreticalUncertaintiesType} we obtain \qCritClaeys, which takes on constant values that depend on the donor stellar type and whether or not the accretor is a degenerate star. For giant stars, they use a function of the core mass ratio \citep{Claeys_etal.2014_TheoreticalUncertaintiesType, Hurley_etal.2002_EvolutionBinaryStars}. The \qCritClaeys prescription is defined based on the initial donor response at the onset of mass loss, and does not capture the delayed stability effects discussed in Sec.~\ref{sec:stability_criteria}. 

The prescriptions from \citet{Ge_etal.2020_AdiabaticMassLoss} use detailed 1D adiabatic mass loss models to account for the evolution of stability over the course of the \ac{MT} episode (see Sec.~\ref{sec:stability_criteria}), and build off of \citet{Ge_etal.2010_AdiabaticMassLoss, Ge_etal.2015_AdiabaticMassLoss} to provide \qCrit as a function of both donor mass and radius (or analagously, evolutionary state).  The parameter \qCritGeTwenty refers to their critical mass ratios calculated using adiabatic mass loss from their standard stellar profiles (including super-adiabatic regions where relevant), while the \qCritGeTwentyIC variant refers to the critical mass ratios calculated when the donor envelopes are artificially isentropic, as a workaround for the rapid super-adiabatic expansion observed in their standard models (see \citealt{Ge_etal.2020_AdiabaticMassLoss} for a more thorough explanation). 

The $\zeta_{ad}$ prescriptions are shown in Fig.~\ref{fig:zetaGrid} as a function of donor radius for a grid of masses. Note that \zetaGeTwenty is calculated directly from the critical mass ratio \qCritGeTwenty, as $\zetaGeTwenty = \zetaRL(\qCritGeTwenty; \beta=1)$, and similarly for $\zetaGeTwentyIC$ and $\zetaClaeys$. We include them in Fig.~\ref{fig:zetaGrid} to highlight differences with \zetaSPH. In practice, the \citet{Ge_etal.2020_AdiabaticMassLoss} values are interpolated in both mass and radius. The vertical lines represent the stellar radii in our models when the stellar type transitions to the named type.

\section{Results}
\label{sec:results}

\subsection{Impact of \texorpdfstring{$\gamma$}{gamma} variations}
\label{sec:impact_gamma}

\begin{figure}[htpb]
\gridline{
  \fig{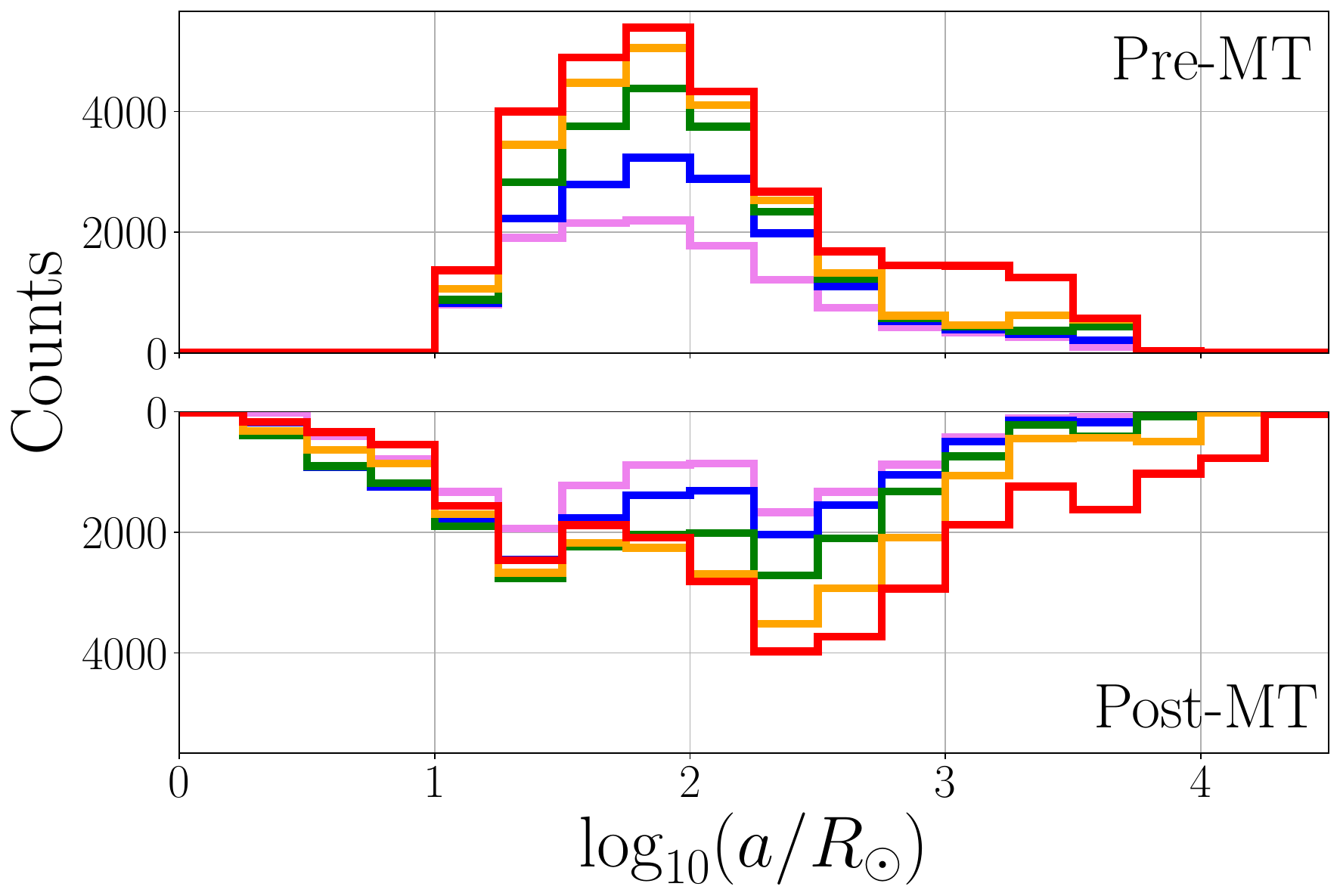}{.95\columnwidth}{(a) Semi-major axis, pre- and post-MT}\label{fig:apre_apost_doublehist}
  }
\gridline{
  \fig{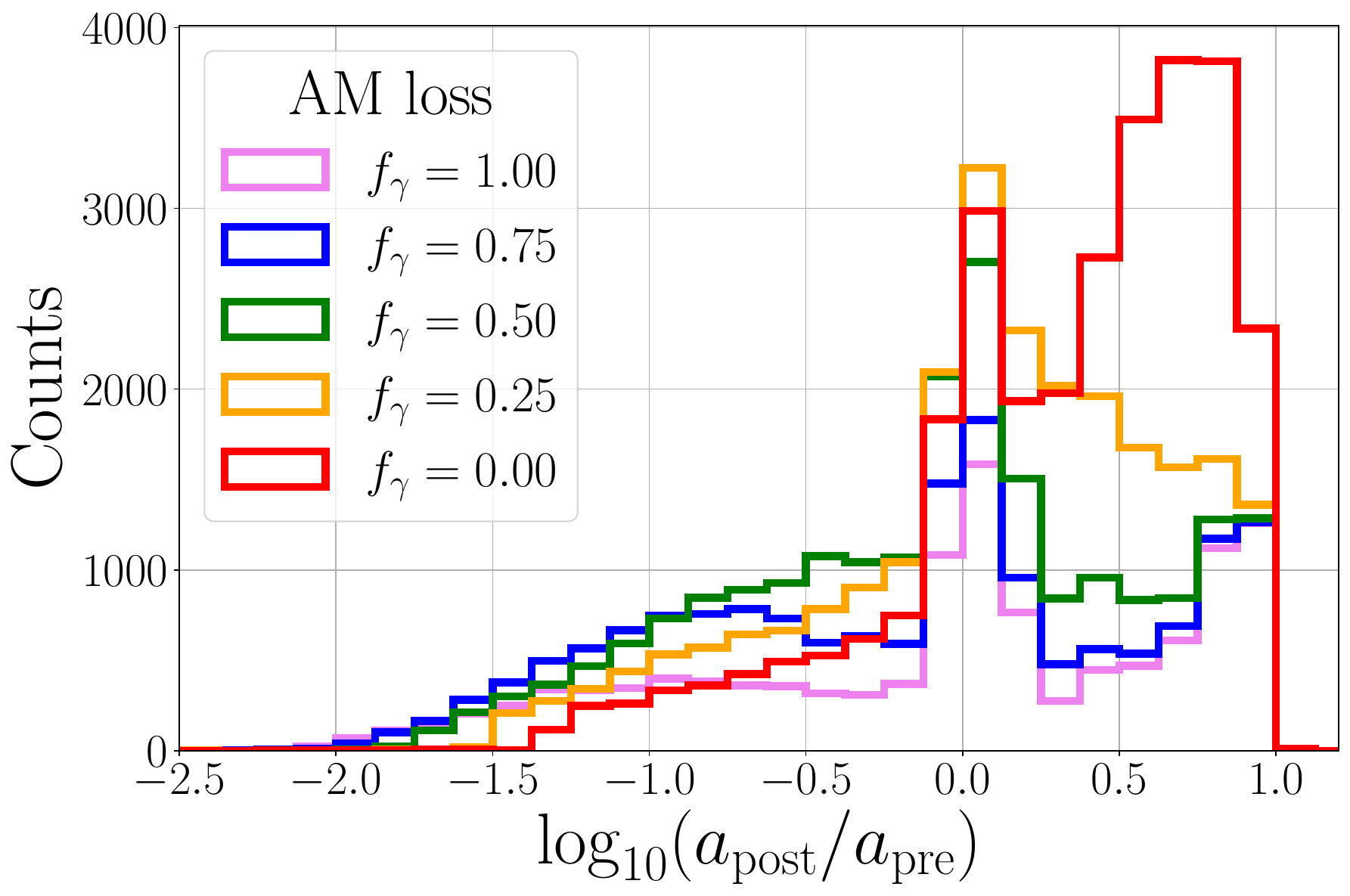}{.95\columnwidth}{(b) Orbital tightening during MT}\label{fig:apre_apost_ratio}
  }
\caption{
\textbf{Distributions of the binary orbital separations before and after the first episode of mass transfer.} 
Panel (a) shows the orbital separations before (pre-\ac{MT}, upper half) and after (post-MT, lower half) the interaction. The post-MT distributions have been inverted. Only systems which undergo stable mass transfer during this first interaction are included. Panel (b) shows the orbital tightening ratios $a_\mathrm{post}/a_\mathrm{pre}$ for the same systems. Both use the default stellar response \zetaSPH and efficiency \betaCompas. Colors in both panels correspond to different values of the specific \ac{AM} loss parameter \fGamma, as indicated in the legend, where $\fGamma=0$ corresponds to isotropic re-emission and $\fGamma=1$ is mass loss from the L2 point (see text for discussion).
}
\label{fig:apre_apost_all}
\end{figure}

For a given binary undergoing \ac{RLOF}, the \ac{MT} efficiency $\beta$ and the specific \ac{AM} of non-accreted material $\gamma$ completely determine the response of the Roche lobe, as well as the final separation of the binary following stable \ac{MT}, via Eq.~(\ref{eq:separation_change}).

In Fig.~\ref{fig:apre_apost_all}, we show how the post-\ac{MT} orbital separations of binaries that experienced stable \ac{MT} depend on the specific \ac{AM} loss through our proxy parameter \fGamma. Here, we follow default assumptions in the other parameters, namely, $\zeta_* = \zetaSPH$ and $\beta=\betaCompas$.

Fig.~\ref{fig:apre_apost_doublehist} shows the separation distribution before (pre-MT, top half) and after (post-MT, bottom half) the first episode of mass transfer. Fig.~\ref{fig:apre_apost_ratio} shows the orbital tightening ratio $\log(\aPostOnAPre)$ for the same systems. Different colors represent different values for \fGamma, as indicated in the legend. Only binaries that undergo stable \ac{MT} are shown, so variations in the normalization for different values of \fGamma correspond to differences in the number of binaries which remain stable.

As expected, increasing \fGamma leads to a reduction in the number of systems that undergo stable \ac{MT}, as many of these are driven toward instability. This is particularly prominent for the pre-MT distributions (upper half of Fig.~\ref{fig:apre_apost_doublehist}), in the range $\log(a/R_\odot) \in \sim[2.7, 3.5]$, which show a population of systems that undergo stable \ac{MT} only for very small values of \fGamma. The post-MT separation distributions cluster into two peaks near $\log(a/R_\odot) \sim$ 1.5 and 2.5. For $\fGamma=0.0$, there is also a prominent bump toward large post-MT separations which is not seen in the pre-MT distribution, indicating that these are systems which ultimately widened as a result of the \ac{MT}.

This can be seen more directly in Fig.~\ref{fig:apre_apost_ratio}, which shows the fractional amount of tightening or widening that binaries experience. For most values of \fGamma shown, the separation ratio distribution is bimodal, with peaks at $\aPostOnAPre\sim1$ (little net change in the separation) and $\aPostOnAPre \sim 5$ (substantial widening). Interestingly, the peak at $\aPostOnAPre\sim 5$ becomes relatively less prominent with decreasing \fGamma, except for $\fGamma=0.00$ when that peak dominates over the one at $\aPostOnAPre\sim1$.

To understand this behavior, we highlight that, for fully conservative mass transfer, the orbital separation decreases until the mass ratio reverses. While this is only approximately true for non-conservative mass transfer, it is nonetheless illustrative. For nearly equal mass ratio binaries undergoing Case A \ac{MT}, $\betaCompas \approx 1$, so the mass ratio will reverse after only a modest amount of mass is lost, and the mass transfer will cease when the Roche lobe encompasses a donor radius only moderately smaller than it was at the onset of \ac{MT}. These binaries thus contribute primarily to the peak at $\aPostOnAPre\sim1$, independently of the \ac{AM} loss prescription. 

Case A \ac{MT} in unequal mass ratio binaries will initially be non-conservative, leading to a dependence on the \ac{AM} loss. However, as the donor mass approaches that of the accretor, the \ac{MT} efficiency $\beta$ will rise towards unity. If the orbit remains sufficiently wide following the initial phase of non-conservative \ac{MT}, the mass ratio will reverse and the the orbit will widen from its minimum value. These binaries thus contribute primarily to the extended tail toward low $\aPostOnAPre$ values. If instead the orbit is too tight following this phase, due perhaps to a high $\fGamma$ value, the two stars will merge. 

Case B systems, by contrast, are nearly always non-conservative, $\betaCompas \to 0$. If the binary has a mass ratio close to unity, the orbit will not shrink much before the mass ratio reversal, and is unlikely to become unstable. The mass loss will then act to substantially widen the binary, populating the second peak at $\aPostOnAPre \sim 5$.

\subsection{Impact of \texorpdfstring{$\gamma$-$\beta$}{gamma-beta} variations} \label{sec:impact_gammabeta}

\begin{figure*}[htpb]
\gridline{
  \fig{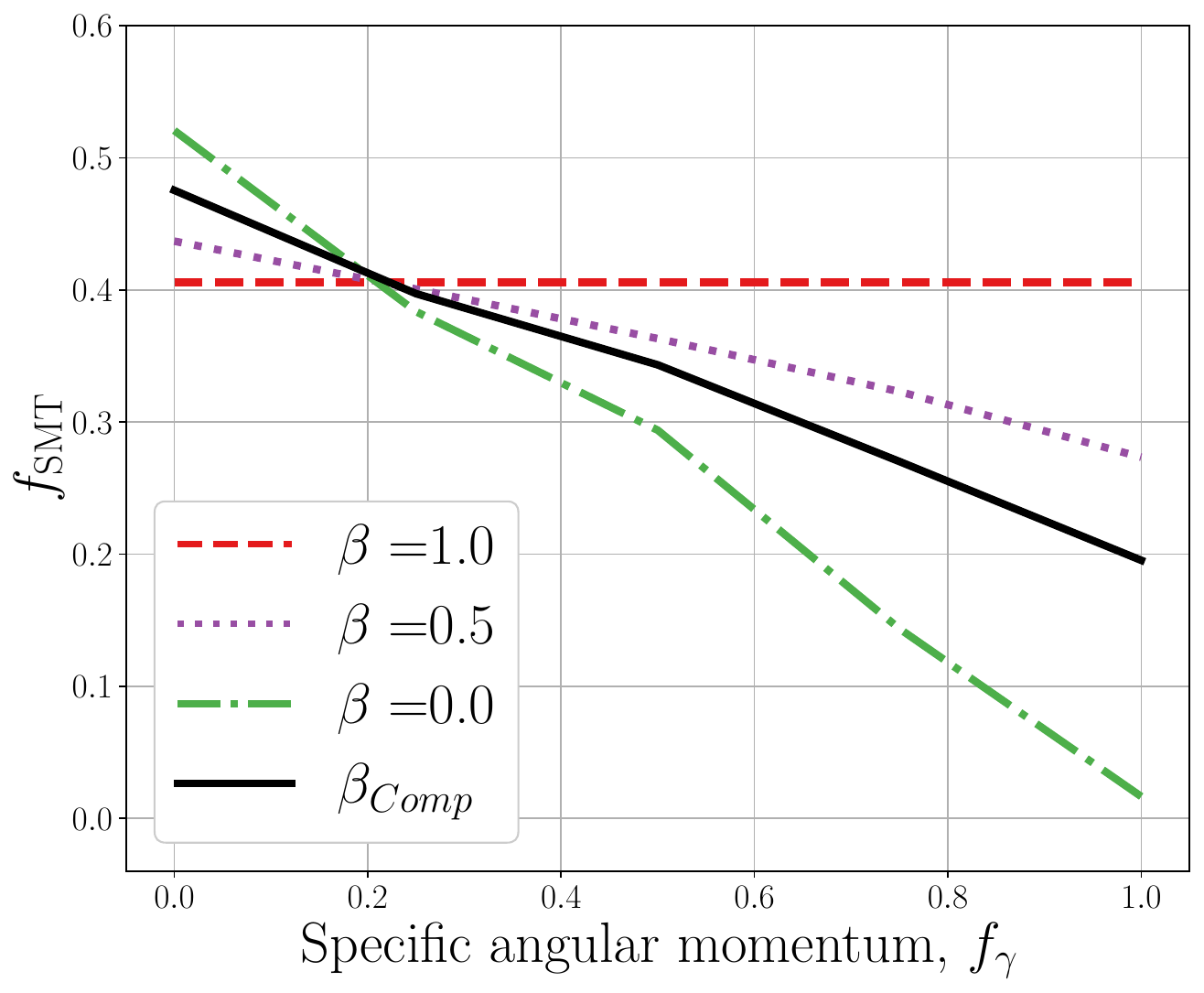}{.95\columnwidth}{(a) SMT fraction after 1st MT}\label{fig:gammaBeta_mt1_uniform}
  \fig{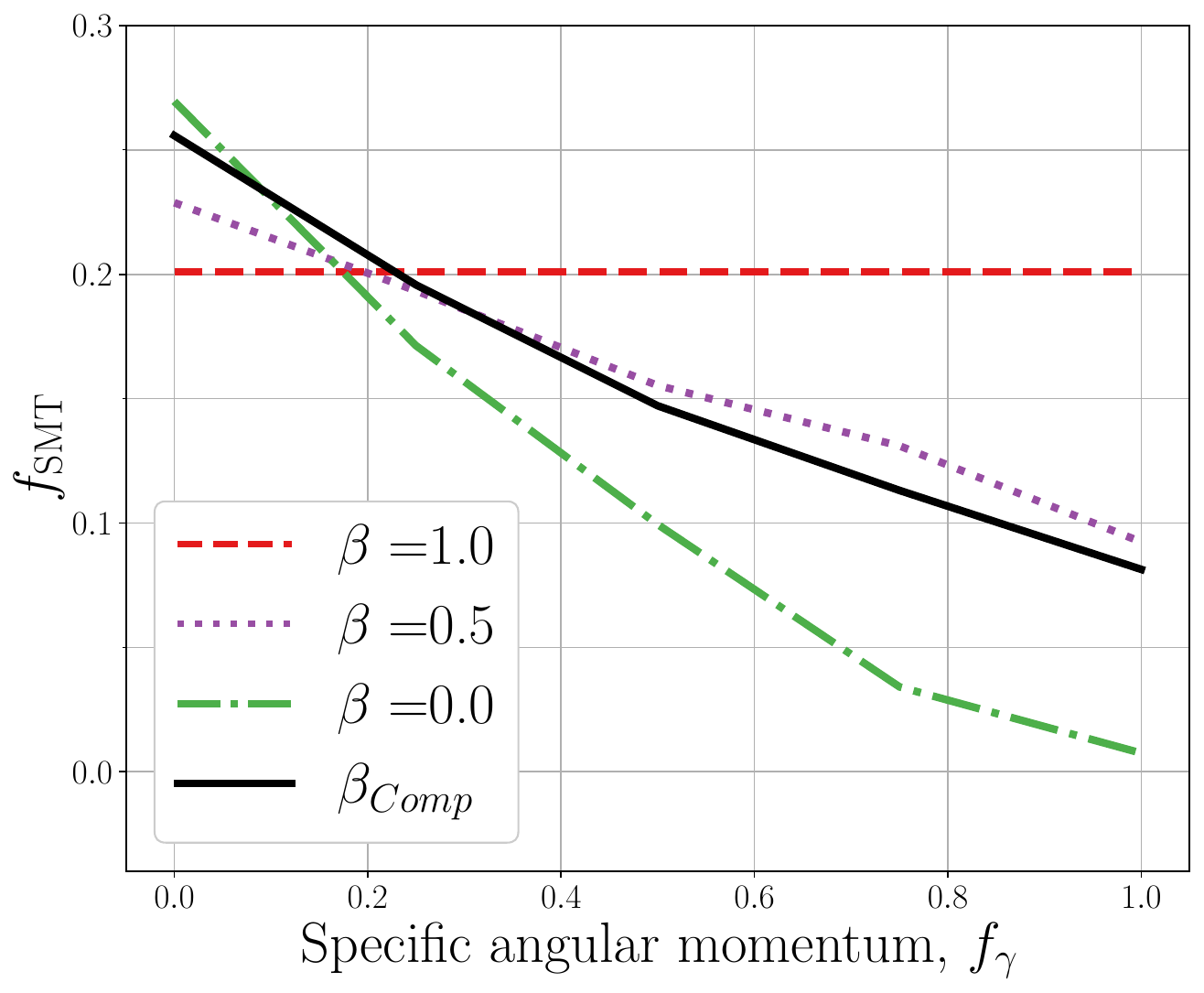}{.95\columnwidth}{(b) SMT fraction at the End state}\label{fig:gammaBeta_end_uniformIC}
  }
\caption{
\textbf{Outcomes of mass transfer, as a function of accretion efficiency and specific angular momentum.} The ordinate shows the fraction of interacting systems which undergo stable mass transfer \fSMT. The remainder (1-\fSMT) experience unstable MT, possibly resulting in a merger. The abscissa indicates different values of the specific \ac{AM} parameter \fGamma, with $\fGamma=0$ corresponding to isotropic re-emission and $\fGamma=1$ corresponding to mass loss from the L2 point. Different curves represent different values of the \ac{MT} efficiency parameter $\beta=\{1.0, 0.5, 0.0\}$, corresponding to red dashed, blue dotted, and green dot-dashed, respectively. The solid black lines show the non-constant COMPAS default $\betaCompas$ (see text for further details). Panel (a) shows \fSMT for the first interaction in a given binary. Panel (b) shows this fraction at the End state (see Sec.~\ref{sec:popsynth}). Here, \fSMT corresponds to the fraction of interacting binaries that have experienced \emph{only} stable \ac{MT} throughout all prior interactions. The donor's response to mass loss is $\zeta_* = \zetaSPH$ for all curves. As expected, \fSMT decreases with increasing \fGamma (if $\beta \neq 1$) at both epochs. Notably, there is a subset of binaries which are only stable if the \ac{MT} is non-conservative, and if $\fGamma \lesssim 0.2$ (as can be seen in Fig.~\ref{fig:zetaRL_heatmap}, an increase in $\beta$ and a decrease in \fGamma both increase the likelihood of stability for a given system).
}
\label{fig:gammaBeta_comparison}
\end{figure*}

The impact of \fGamma variations is maximal for low values of the \ac{MT} efficiency $\beta$ and reduced as $\beta \to 1$.  We investigate the influence of $\beta$ and \fGamma by varying both parameters simultaneously in Fig.~\ref{fig:gammaBeta_comparison}.  In Fig.~\ref{fig:gammaBeta_mt1_uniform}, we show how the stable \ac{MT} fraction during the first \ac{MT} episode depends on the adopted value of \fGamma (shown on the abscissa), assuming the default donor response $\zeta_* = \zetaSPH$. Higher $\beta$ values (represented by line color) decrease the sensitivity of this \fGamma dependence. Fig.~\ref{fig:gammaBeta_end_uniformIC} shows the same analysis for systems evaluated prior to the first \ac{SN}, which we discuss in more detail in Sec.~\ref{sec:long_term}. 

The height of each line represents the stable \ac{MT} fraction \fSMT. The remaining fraction, $1-\fSMT$, includes systems that experienced unstable \ac{MT} during the first interaction, including any stellar mergers. Non-interacting systems are excluded. When $\beta=0$, \fSMT varies over a very broad range from $\sim0-50\%$. Even for more realistic values of $\beta=\betaCompas$, the range is still substantial, closer to $\sim20-50\%$.

\subsection{Donor radial response \texorpdfstring{$\zeta_{ad}$ and \qCrit}{ZetaAd and qCrit}}
\label{sec:impact_donor_radius}

\begin{figure*}[!htpb]
\centering
\includegraphics[width=.8\paperwidth]{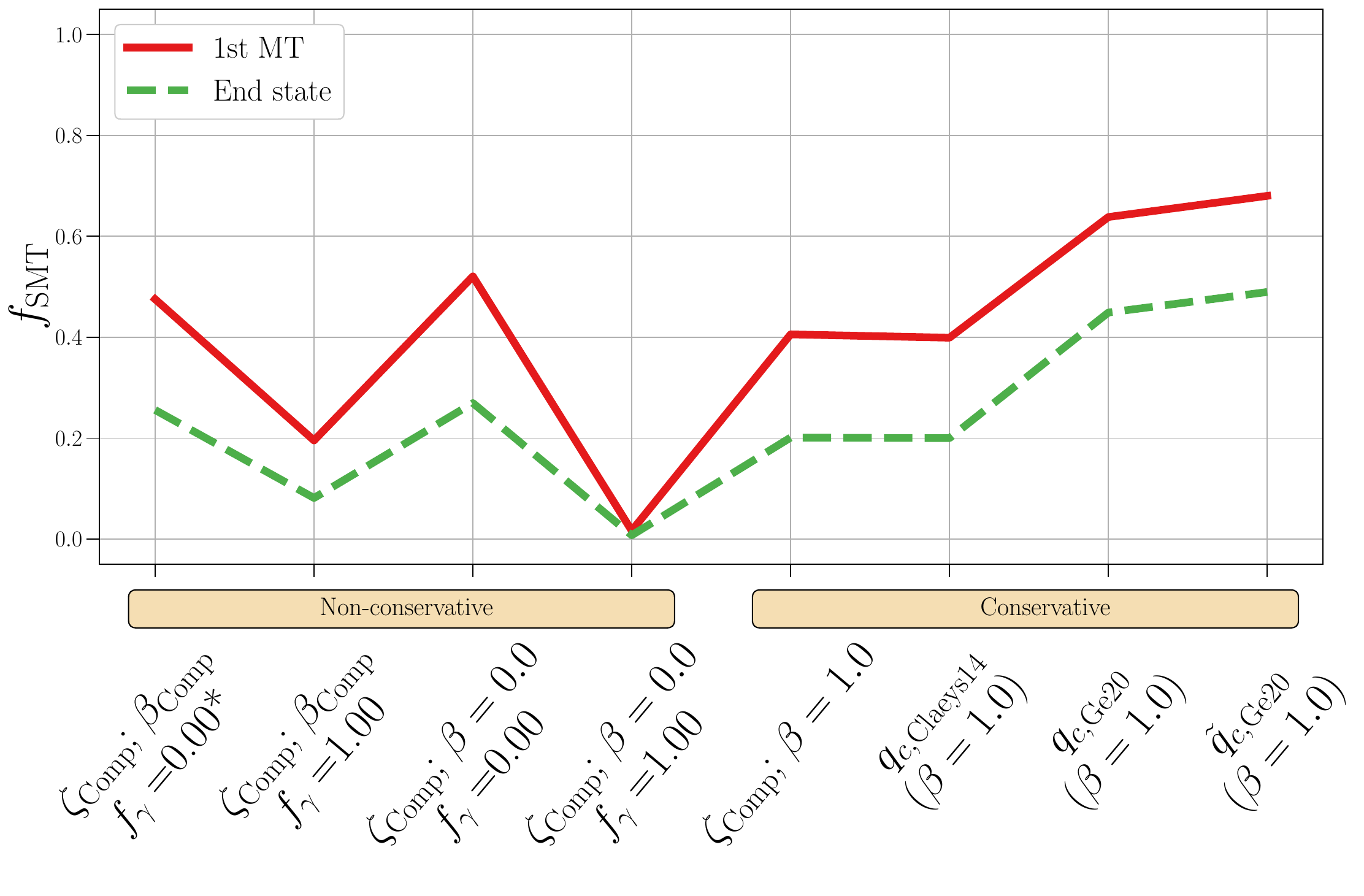}
\caption{
\textbf{Stable mass transfer fraction, for all model variations.}
As in Fig.~\ref{fig:gammaBeta_comparison}, we plot the fraction of interacting systems which experienced only stable mass transfer \fSMT during the first \ac{MT} event (solid, red curve), and by the End state (green, dashed curve), see Sec.~\ref{sec:popsynth}. Model names are listed on the abscissa. The left-most starred model is the COMPAS default. Models on the left fixed the donor radial response to mass loss \zetaSPH while varying the accretion parameters $\beta$ and \fGamma. Those on the right varied the donor response via the stability condition (assuming fully conservative \ac{MT}). 
}
\label{fig:mostModsComparison}
\end{figure*}

In Fig.~\ref{fig:mostModsComparison}, we show \fSMT across all model variations following the first \ac{MT} episode (solid, red line) and at the End state (dashed, green line), which we discuss further in Sec.~\ref{sec:long_term}. Model variations are listed on the abscissa. The ordering groups together similar models; those on the left fix $\zeta=\zetaSPH$ and vary $\beta$ and $\gamma$, while those on the right vary the donor response prescription and implicitly fix $\beta=1$. For models which depend on $\beta$ and \fGamma, we include only the extremal values of these parameters (as well as our default \betaCompas) for brevity. 

All of the models which define stability based on critical mass ratios \qCrit assume fully conservative mass transfer ($\beta=1$). The \zetaSPH model furthest to the right also includes fully conservative mass transfer to facilitate comparison with the critical mass ratio models. The (\zetaSPH, $\beta=1$) model and \qCritClaeys both define stability at the onset of \ac{MT}, i.e. ignoring delayed instabilities. 

By contrast, the \citet{Ge_etal.2020_AdiabaticMassLoss} models consider the evolution of stability over the course of \ac{MT}. The aggregate effect is to increase the fraction \fSMT of systems undergoing stable \ac{MT} during the 1st \ac{MT} event from $\sim 40\%$ for the (\zetaSPH, $\beta=1$) model to $\sim 70\%$ for the \citet{Ge_etal.2020_AdiabaticMassLoss} models. This is primarily due to the reduction in convective envelope donors which experience instability in \citet{Ge_etal.2020_AdiabaticMassLoss}, since the delayed dynamical instability mechanism in radiative donors acts in the other direction, to increase the number of unstable systems that would previously have been considered stable. 

Meanwhile, we see as before that varying just $\beta$ and $\gamma$ at fixed $\zeta_* = \zetaSPH$ drives significant changes in \fSMT. The maximal variation in \fSMT from all variations in $\beta$ and $\gamma$ is $\sim50\%$ -- seen in the difference between (\zetaSPH, $\beta=0.0$, $\fGamma=0.0$) and (\zetaSPH, $\beta=0.0$, $\fGamma=1.0$). For the default COMPAS efficiency $\beta=\betaCompas$, \fSMT varies by $\sim25\%$ across the full range of \fGamma. Fluctuations in the stable mass transfer fraction \fSMT due to changes in the efficiency and specific \ac{AM} loss during mass transfer are comparable in magnitude to the variations from the donor stability prescriptions.

\subsection{Mass transfer outcomes pre-supernova}
\label{sec:long_term}

In Fig.~\ref{fig:gammaBeta_end_uniformIC}, we consider the impacts of the $\beta-\gamma$ variations at the End State. We find that $\sim0-20\%$ of interacting binaries will undergo only stable \ac{MT} throughout their evolution.

As with the first \ac{MT} episode, the stable mass transfer fraction at the End state is strongly dependent on the adopted \ac{AM} loss prescription \fGamma (for values of $\beta<1$), but here the impact is more concentrated at low \fGamma values. For $\fGamma\gtrsim0.7$, $\fSMT$ drops to a few percent with only a weak dependence on \fGamma, which suggests that if \fGamma is universally high, even binaries which remain stable during the first \ac{MT} will eventually experience unstable \ac{MT}. 

We show \fSMT at the End state across all model variations in Fig.~\ref{fig:mostModsComparison} (dashed, green line). Here, the differences between models are qualitatively similar to differences following the first \ac{MT} episode, though not quite as pronounced. Interestingly, although \fSMT is still sensitive to $\beta-\gamma$ variations at the end of the evolution, the impact of different donor responses (for fully conservative \ac{MT}, $\beta=1$) is greater at this epoch, with \fSMT ranging from $\sim15-45\%$. Taken at face value, this suggests that uncertainties in the donor response are more influential for the long term evolution than uncertainties in the Roche-lobe response, though we emphasize that it is difficult to estimate without models that vary both the donor response and the Roche-lobe response simultaneously.

\section{Discussion and Conclusion}
\label{sec:discussion_conclusion}

We investigated the conditions and outcomes of stable mass transfer in populations of interacting, isolated binaries. Model variations included the treatment of transferred material and its impact on the Roche lobe, and the radial response of the donor to mass loss. We focused in particular on the fraction of systems that experience stable \ac{MT} during the first \ac{MT} event and the effect on the post-\ac{MT} orbital separations, as well as the fraction of systems that experience only stable \ac{MT} events throughout their evolution, prior to any supernovae.

\subsection{Sensitivity to \texorpdfstring{$\gamma$}{gamma}}
\label{sec:sensitivity_to_gamma}

The fraction of systems that undergo stable \ac{MT} during the first \ac{MT} event, as well as the fraction of systems that will experience only stable \ac{MT} throughout their evolution, strongly depends on the specific angular momentum $\gamma$ carried away by non-accreted material (or equivalently, our proxy parameter \fGamma), as shown in Fig.~\ref{fig:mostModsComparison}. \ac{BPS} codes often assume that non-accreted matter follows the ``isotropic re-emission'' model, taking with it the the specific \ac{AM} of the accretor, i.e. $\fGamma=0$, in all cases. However, both models and observations suggest that this may not be the case. 

Using detailed hydrodynamic simulations, \citet{MacLeod_etal.2018_BoundOutflowsUnbound, MacLeod_etal.2018_RunawayCoalescenceOnset} find that $\fGamma$ should not be constant, but depends sensitively on properties of the system, such as the evolutionary state of the donor. \citet{Kalogera_Webbink.1996_FormationLowMassXRay} argue that $\fGamma\sim0$ if the accretor is an \ac{NS} or \ac{BH} on the basis that gas ejection is driven by accretion energy deposited very close to the compact object. However, \citet{Goodwin_Woods.2020_BinaryEvolutionSAX} find that $\fGamma < 0$ (mass lost near the L1 point) is required to explain the large period derivative of SAX J1808.4-3658. Moreover, there may exist mechanisms to keep ejected matter synchronized to the binary orbit beyond the L2 point, such as magnetic fields, collisions in the ejecta streams, or continuous torquing from circumbinary disks. Thus \fGamma may not be \emph{a priori} limited to $\fGamma \leq 1$.

We find that the fraction of systems which undergo stable \ac{MT} during the first \ac{MT} event is sensitive to the value of \fGamma across the entire range $\fGamma\in[0,1]$, but that by the end of the evolution, nearly all systems experience \ac{CEE} at some point if $\fGamma \gtrsim 0.7$ (see Fig.~\ref{fig:gammaBeta_comparison}). If very high $\fGamma$ values are more representative of reality, this would efficiently kill off the stable \ac{MT} channel for the formation of binaries containing a compact object (in agreement with \citealt{vanSon_etal.2022_NoPeaksValleys}, in the context of binary black hole formation).

\subsection{Uncertainties in donor response}
\label{sec:uncertainties_in_donor_response}

Mass transfer stability prescriptions based on the delayed onset of mass transfer instability in both radiative and convective envelopes, such as those from \citet{Ge_etal.2020_AdiabaticMassLoss}, show that stable mass transfer is significantly more common than is suggested by prescriptions that are restricted to the onset of mass transfer. However, these more nuanced prescriptions are computed based on the assumption of fully conservative \ac{MT}. To the extent that this is not an accurate representation of most binary interactions, this assumption under-predicts the likelihood of instability.

In Fig.~\ref{fig:zetaGrid}, we show the prescriptions for $\zeta_{ad}$ derived from the detailed models from \citet{Ge_etal.2020_AdiabaticMassLoss}, \zetaGeTwenty and \zetaGeTwentyIC, as well as the default \zetaSPH used in COMPAS, and \zetaClaeys from \citet{Claeys_etal.2014_TheoreticalUncertaintiesType}. \zetaGeTwenty and \zetaGeTwentyIC are generally very similar, but both are higher than \zetaSPH and \zetaClaeys for virtually all masses and radii, indicating that stable \ac{MT} may be more abundant than is currently estimated, variations to $\beta$ and $\gamma$ notwithstanding. 

Notably, in response to \citet{Ge_etal.2010_AdiabaticMassLoss}, \citet{Woods_Ivanova.2011_CanWeTrust} argue that the thermal timescale in the super-adiabatic surface layers is comparable to the dynamical timescale, thus invalidating the adiabatic assumption of \citet{Ge_etal.2010_AdiabaticMassLoss, Ge_etal.2015_AdiabaticMassLoss, Ge_etal.2020_AdiabaticMassLoss}. They predict that the thermal contraction and restructuring of the superadiabatic layer occurs faster than the mass can be removed, except for unrealistically high mass loss rates. This prediction is supported by \citet{Pavlovskii_Ivanova.2015_MassTransferGiant}, who find no evidence for rapid expansion when recombination energy is included in the superadiabatic layers. They argue instead that matter flowing through the nozzle around the L1 point is constricted, and that \ac{MT} instability only sets in once stellar material begins to overflow through either the L2 or L3 points \citep{Pavlovskii_Ivanova.2015_MassTransferGiant, Pavlovskii_etal.2017_StabilityMassTransfer}. \citet{Lu_etal.2022_RapidBinaryMass} explored the threshold conditions for mass loss from the L2 point, and find that this occurs only for fairly high donor mass loss rates $\gtrsim 10^{-4}\;\Msun/\mathrm{yr}$, though this is dependent on the geometry of the system and the assumed cooling rate of the accretion stream. This is marginally consistent with \citet{Ge_etal.2010_AdiabaticMassLoss}, who find that \ac{MT} instability sets in around $\sim 10^{-5}\;\Msun/\mathrm{yr}$ but quickly rises above $\sim 10^{-4}\;\Msun/\mathrm{yr}$ (see their Fig. 5), though a full population synthesis study to properly compare the models is warranted.

In a recent paper, \citet{Temmink_etal.2022_CopingLossStability} compared several different stability definitions, including a critical mass loss rate at which the donor response becomes adiabatic, overflow through the L2 point, and rapid contraction of the orbit. They found that \ac{MT} is generally more stable than is predicted in the prescriptions commonly in use in \ac{BPS} codes, although they also computed their stability criteria assuming fully conservative \ac{MT}, and their study focused on a lower mass regime than is considered here. Crucially, they conclude that none of the definitions for stability are self-consistent -- or particularly reliable -- for very evolved stars, which suggests that the difference between stable and unstable \ac{MT} itself may not be well-defined, or perhaps that the transition between the two regimes is not as discrete as has been assumed historically. 

In practice, the convective/radiative nature of an envelope is not directly aligned with its stellar type as defined in \citet{Hurley_etal.2000_ComprehensiveAnalyticFormulae}. Convective instabilities develop in cool envelopes where the opacity is sufficiently high to make radiation transport inefficient. We implemented an improved prescription in which a giant's envelope is radiative if the surface effective temperature is greater than a given threshold temperature, and convective otherwise. We considered several values for this threshold temperature, including $\log(T_\mathrm{eff}/\mathrm{K}) = 3.73$ from \citet{Belczynski_etal.2008_CompactObjectModeling} and $\log(T_\mathrm{eff}/\mathrm{K}) = 3.65$ from \citet{Klencki_etal.2021_ItHasBe}. However, we found that these variations had only a marginal impact on mass transfer stability, and thus did not pursue them further. This may suggest that the convective/radiative boundary is not so uncertain as to substantially impact the stable mass transfer fraction, however population synthesis with more detailed stellar structures and responses to mass loss will certainly provide some clarity here.

\subsection{Observational constraints}
\label{sec:observables}

Observations of interacting binaries are crucial in constraining the parameters $\beta$ and $\gamma$. Binaries engaged in ongoing \ac{MT} can provide vital constraints on the instantaneous values of $\beta$ and $\gamma$ in a given mass exchange episode, via measurable changes to the orbital period. In tandem, the properties of observed populations of specific classes of post-interaction binaries, or their associated transients, can also be informative in determining the values and universality of $\beta$ and $\gamma$ in different contexts, though these are subject to modelling constraints.

For an in-depth review of the possible classes of interacting binary observations, we refer the reader to \citet{DeMarco_Izzard.2017_DawesReviewImpact}, particularly their Table~1 and the references therein. However, we emphasize that the most useful observations will be those which involve binaries which have experienced stable mass transfer recently in their evolution (e.g. Algols), to reduce modelling uncertainties. By contrast, binaries which are believed to have survived a supernova or common envelope event, must necessarily be convolved with models for these more uncertain phases, reducing the viability of any claimed constraints. The choice taken in this paper to study the stable mass transfer rate only up to the first supernova reflects this preference to reduce the modelling complexity in order to derive more robust conclusions.

\section*{Acknowledgements}

We thank Rosanne Di Stefano for productive discussions on the limitations of current observations of multiple stars, Max Moe and Mads S\o rensen for providing the correlated distribution sampler (see App. \ref{sec:impact_correlated_ICs}), and Michela Mapelli, Fabian Schneider, Floor Broekgaarden, and members of Team COMPAS for useful feedback on the manuscript. RTW thanks the Harvard Center for Astrophysics for its hospitality while this work was completed. RTW, IM, and RH are supported by the Australian Research Council Centre of Excellence for Gravitational Wave Discovery (OzGrav), through project number CE170100004.  This work made use of the \mbox{OzSTAR} high performance computer at Swinburne University of Technology.  \mbox{OzSTAR} is funded by Swinburne University of Technology and the National Collaborative Research Infrastructure Strategy (NCRIS). IM is a recipient of the Australian Research Council Future Fellowships (FT190100574). MM gratefully acknowledges support of the Clay postdoctoral fellowship of the Smithsonian Astrophysical Observatory.

\section*{Code availability}

Simulations in this paper made use of COMPAS \citep{TeamCOMPAS_etal.2022_COMPASRapidBinary} v02.38.07. Data analysis was performed using python v3.9 \citep{vanRossum.1995_PythonTutorial}, numpy v1.21 \citep{Harris_etal.2020_ArrayProgrammingNumPy}, and matplotlib v3.5 \citep{Hunter.2007_Matplotlib2DGraphics}. Data and post-processing scripts are available on request to the corresponding author.

\bibliographystyle{aasjournal}
\bibliography{bib.bib}

\begin{thebibliography}{}
\expandafter\ifx\csname natexlab\endcsname\relax\def\natexlab#1{#1}\fi
\providecommand{\url}[1]{\href{#1}{#1}}
\providecommand{\dodoi}[1]{doi:~\href{http://doi.org/#1}{\nolinkurl{#1}}}
\providecommand{\doeprint}[1]{\href{http://ascl.net/#1}{\nolinkurl{http://ascl.net/#1}}}
\providecommand{\doarXiv}[1]{\href{https://arxiv.org/abs/#1}{\nolinkurl{https://arxiv.org/abs/#1}}}

\bibitem[{Belczynski {et~al.}(2008)Belczynski, Kalogera, Rasio, Taam, Zezas,
  Bulik, Maccarone, \& Ivanova}]{Belczynski_etal.2008_CompactObjectModeling}
Belczynski, K., Kalogera, V., Rasio, F.~A., {et~al.} 2008, The Astrophysical
  Journal Supplement Series, 174, 223, \dodoi{10.1086/521026}

\bibitem[{Claeys {et~al.}(2014)Claeys, Pols, Izzard, Vink, \&
  Verbunt}]{Claeys_etal.2014_TheoreticalUncertaintiesType}
Claeys, J. S.~W., Pols, O.~R., Izzard, R.~G., Vink, J., \& Verbunt, F. W.~M.
  2014, Astronomy \& Astrophysics, 563, A83,
  \dodoi{10.1051/0004-6361/201322714}

\bibitem[{COMPAS {et~al.}(2022)COMPAS, Riley, Agrawal, Barrett, Boyett,
  Broekgaarden, Chattopadhyay, Gaebel, Gittins, Hirai, Howitt, Justham,
  Khandelwal, Kummer, Lau, Mandel, {de Mink}, Neijssel, Riley, {van Son},
  Stevenson, {Vigna-G{\'o}mez}, Vinciguerra, Wagg, \&
  Willcox}]{TeamCOMPAS_etal.2022_COMPASRapidBinary}
COMPAS, T., Riley, J., Agrawal, P., {et~al.} 2022, The Journal of Open Source
  Software, 7, 3838, \dodoi{10.21105/joss.03838}

\bibitem[{De~Marco \& Izzard(2017)}]{DeMarco_Izzard.2017_DawesReviewImpact}
De~Marco, O., \& Izzard, R.~G. 2017, Publications of the Astronomical Society
  of Australia, 34, e001, \dodoi{10.1017/pasa.2016.52}

\bibitem[{{de Mink} {et~al.}(2007){de Mink}, Pols, \&
  Hilditch}]{deMink_etal.2007_EfficiencyMassTransfer}
{de Mink}, S.~E., Pols, O.~R., \& Hilditch, R.~W. 2007, Astronomy and
  Astrophysics, 467, 1181, \dodoi{10.1051/0004-6361:20067007}

\bibitem[{Dosopoulou \&
  Kalogera(2018)}]{Dosopoulou_Kalogera.2018_OrbitalEvolutionMassTransferring}
Dosopoulou, F., \& Kalogera, V. 2018

\bibitem[{Eggleton(1983)}]{Eggleton.1983_AproximationsRadiiRoche}
Eggleton, P.~P. 1983, The Astrophysical Journal, 268, 368,
  \dodoi{10.1086/160960}

\bibitem[{Ge {et~al.}(2010)Ge, Hjellming, Webbink, Chen, \&
  Han}]{Ge_etal.2010_AdiabaticMassLoss}
Ge, H., Hjellming, M.~S., Webbink, R.~F., Chen, X., \& Han, Z. 2010, The
  Astrophysical Journal, 717, 724, \dodoi{10.1088/0004-637X/717/2/724}

\bibitem[{Ge {et~al.}(2015)Ge, Webbink, Chen, \&
  Han}]{Ge_etal.2015_AdiabaticMassLoss}
Ge, H., Webbink, R.~F., Chen, X., \& Han, Z. 2015, The Astrophysical Journal,
  812, 40, \dodoi{10.1088/0004-637X/812/1/40}

\bibitem[{Ge {et~al.}(2020)Ge, Webbink, Chen, \&
  Han}]{Ge_etal.2020_AdiabaticMassLoss}
---. 2020, The Astrophysical Journal, 899, 132,
  \dodoi{10.3847/1538-4357/aba7b7}

\bibitem[{Goodwin \& Woods(2020)}]{Goodwin_Woods.2020_BinaryEvolutionSAX}
Goodwin, A.~J., \& Woods, T.~E. 2020, Monthly Notices of the Royal Astronomical
  Society, 495, 796, \dodoi{10.1093/mnras/staa1234}

\bibitem[{Grudi{\'c} {et~al.}(2023)Grudi{\'c}, Offner, Guszejnov,
  {Faucher-Gigu{\`e}re}, \& Hopkins}]{Grudic_etal.2023_DoesGodPlay}
Grudi{\'c}, M.~Y., Offner, S. S.~R., Guszejnov, D., {Faucher-Gigu{\`e}re},
  C.-A., \& Hopkins, P.~F. 2023, Does {{God}} Play Dice with Star Clusters?,
  {arXiv}.
\newblock \doeprint{2307.00052}

\bibitem[{Harris {et~al.}(2020)Harris, Millman, {van der Walt}, Gommers,
  Virtanen, Cournapeau, Wieser, Taylor, Berg, Smith, Kern, Picus, Hoyer, {van
  Kerkwijk}, Brett, Haldane, {del R{\'i}o}, Wiebe, Peterson,
  {G{\'e}rard-Marchant}, Sheppard, Reddy, Weckesser, Abbasi, Gohlke, \&
  Oliphant}]{Harris_etal.2020_ArrayProgrammingNumPy}
Harris, C.~R., Millman, K.~J., {van der Walt}, S.~J., {et~al.} 2020, Nature,
  585, 357, \dodoi{10.1038/s41586-020-2649-2}

\bibitem[{Hirai \&
  Mandel(2022)}]{Hirai_Mandel.2022_TwostageFormalismCommonenvelope}
Hirai, R., \& Mandel, I. 2022, A Two-Stage Formalism for Common-Envelope Phases
  of Massive Stars

\bibitem[{Hjellming \&
  Webbink(1987)}]{Hjellming_Webbink.1987_ThresholdsRapidMass}
Hjellming, M.~S., \& Webbink, R.~F. 1987, The Astrophysical Journal, 318, 794,
  \dodoi{10.1086/165412}

\bibitem[{Hunter(2007)}]{Hunter.2007_Matplotlib2DGraphics}
Hunter, J.~D. 2007, Computing in Science and Engineering, 9, 90,
  \dodoi{10.1109/MCSE.2007.55}

\bibitem[{Hurley {et~al.}(2000)Hurley, Pols, \&
  Tout}]{Hurley_etal.2000_ComprehensiveAnalyticFormulae}
Hurley, J.~R., Pols, O.~R., \& Tout, C.~A. 2000, Mon. Not. R. Astron. Soc, 000,
  1

\bibitem[{Hurley {et~al.}(2002)Hurley, Tout, \&
  Pols}]{Hurley_etal.2002_EvolutionBinaryStars}
Hurley, J.~R., Tout, C.~A., \& Pols, O.~R. 2002, Monthly Notices of the Royal
  Astronomical Society, 329, 897, \dodoi{10.1046/j.1365-8711.2002.05038.x}

\bibitem[{Ivanova(2017)}]{Ivanova.2017_CommonEnvelopeProgress}
Ivanova, N. 2017, 329, 199, \dodoi{10.1017/S1743921317003398}

\bibitem[{Ivanova \& Nandez(2016)}]{Ivanova_Nandez.2016_CommonEnvelopeEvents}
Ivanova, N., \& Nandez, J. L.~A. 2016, Monthly Notices of the Royal
  Astronomical Society, 462, 362, \dodoi{10.1093/mnras/stw1676}

\bibitem[{Kalogera \&
  Webbink(1996)}]{Kalogera_Webbink.1996_FormationLowMassXRay}
Kalogera, V., \& Webbink, R.~F. 1996, The Astrophysical Journal, 458, 301,
  \dodoi{10.1086/176813}

\bibitem[{Kippenhahn {et~al.}(2012)Kippenhahn, Weigert, \&
  Weiss}]{Kippenhahn_etal.2012_StellarStructureEvolution}
Kippenhahn, R., Weigert, A., \& Weiss, A. 2012, Stellar {{Structure}} and
  {{Evolution}} ({Springer-Verlag Berlin Heidelberg})

\bibitem[{Klencki {et~al.}(2021)Klencki, Nelemans, Istrate, \&
  Chruslinska}]{Klencki_etal.2021_ItHasBe}
Klencki, J., Nelemans, G., Istrate, A.~G., \& Chruslinska, M. 2021, Astronomy
  and Astrophysics, 645, A54, \dodoi{10.1051/0004-6361/202038707}

\bibitem[{Kroupa(2001)}]{Kroupa.2001_VariationInitialMass}
Kroupa, P. 2001, Monthly Notices of the Royal Astronomical Society, 322, 231,
  \dodoi{10.1046/j.1365-8711.2001.04022.x}

\bibitem[{Lau {et~al.}(2022)Lau, Hirai, Price, \&
  Mandel}]{Lau_etal.2022_CommonEnvelopesMassive}
Lau, M. Y.~M., Hirai, R., Price, D.~J., \& Mandel, I. 2022, Monthly Notices of
  the Royal Astronomical Society, \dodoi{10.1093/mnras/stac2490}

\bibitem[{Lu {et~al.}(2022)Lu, Fuller, Quataert, \&
  Bonnerot}]{Lu_etal.2022_RapidBinaryMass}
Lu, W., Fuller, J., Quataert, E., \& Bonnerot, C. 2022, arXiv:2204.00847
  [astro-ph].
\newblock \doeprint{2204.00847}

\bibitem[{MacLeod \&
  Loeb(2020{\natexlab{a}})}]{MacLeod_Loeb.2020_PrecommonenvelopeMassLoss}
MacLeod, M., \& Loeb, A. 2020{\natexlab{a}}, The Astrophysical Journal, 895,
  29, \dodoi{10.3847/1538-4357/ab89b6}

\bibitem[{MacLeod \&
  Loeb(2020{\natexlab{b}})}]{MacLeod_Loeb.2020_RunawayCoalescencePrecommonenvelope}
---. 2020{\natexlab{b}}, The Astrophysical Journal, 893, 106,
  \dodoi{10.3847/1538-4357/ab822e}

\bibitem[{MacLeod {et~al.}(2018{\natexlab{a}})MacLeod, Ostriker, \&
  Stone}]{MacLeod_etal.2018_BoundOutflowsUnbound}
MacLeod, M., Ostriker, E.~C., \& Stone, J.~M. 2018{\natexlab{a}}, The
  Astrophysical Journal, 868, 136, \dodoi{10.3847/1538-4357/aae9eb}

\bibitem[{MacLeod {et~al.}(2018{\natexlab{b}})MacLeod, Ostriker, \&
  Stone}]{MacLeod_etal.2018_RunawayCoalescenceOnset}
---. 2018{\natexlab{b}}, The Astrophysical Journal, 863, 5,
  \dodoi{10.3847/1538-4357/aacf08}

\bibitem[{Marchant {et~al.}(2021)Marchant, Pappas, {Gallegos-Garcia}, Berry,
  Taam, Kalogera, \& Podsiadlowski}]{Marchant_etal.2021_RoleMassTransfer}
Marchant, P., Pappas, K. M.~W., {Gallegos-Garcia}, M., {et~al.} 2021, Astronomy
  and Astrophysics, 650, A107, \dodoi{10.1051/0004-6361/202039992}

\bibitem[{Moe \& Di~Stefano(2017)}]{Moe_DiStefano.2017_MindYourPs}
Moe, M., \& Di~Stefano, R. 2017, The Astrophysical Journal Supplement Series,
  230, 15, \dodoi{10.3847/1538-4365/aa6fb6}

\bibitem[{Neo {et~al.}(1977)Neo, Miyaji, Nomoto, \&
  Sugimoto}]{Neo_etal.1977_EffectRapidMass}
Neo, S., Miyaji, S., Nomoto, K., \& Sugimoto, D. 1977, Publications of the
  Astronomical Society of Japan, 29, 249

\bibitem[{Paczynski(1976)}]{Paczynski.1976_CommonEnvelopeBinaries}
Paczynski, B. 1976, in Structure and {{Evolution}} of {{Close Binary Systems}};
  {{Proceedings}} of the {{Symposium}}, Vol.~73 ({D. Reidel Publishing Co.}),
  75

\bibitem[{Paczy{\'n}ski \&
  Sienkiewicz(1972)}]{Paczynski_Sienkiewicz.1972_EvolutionCloseBinaries}
Paczy{\'n}ski, B., \& Sienkiewicz, R. 1972, Acta Astronomica, 22, 73

\bibitem[{Pavlovskii \&
  Ivanova(2015)}]{Pavlovskii_Ivanova.2015_MassTransferGiant}
Pavlovskii, K., \& Ivanova, N. 2015, Monthly Notices of the Royal Astronomical
  Society, 449, 4415, \dodoi{10.1093/mnras/stv619}

\bibitem[{Pavlovskii {et~al.}(2017)Pavlovskii, Ivanova, Belczynski, \&
  Van}]{Pavlovskii_etal.2017_StabilityMassTransfer}
Pavlovskii, K., Ivanova, N., Belczynski, K., \& Van, K.~X. 2017, Monthly
  Notices of the Royal Astronomical Society, 465, 2092,
  \dodoi{10.1093/mnras/stw2786}

\bibitem[{Pols(2018)}]{Pols.2018_CourseNotesBinary}
Pols, O.~R. 2018, Course {{Notes}} - {{Binary Stars}}.
  {{https://www.astro.ru.nl/\textasciitilde onnop/}}

\bibitem[{Sana {et~al.}(2012)Sana, De~Mink, De~Koter, Langer, Evans, Gieles,
  Gosset, Izzard, Le~Bouquin, \&
  Schneider}]{Sana_etal.2012_BinaryInteractionDominates}
Sana, H., De~Mink, S.~E., De~Koter, A., {et~al.} 2012, Science, 337, 444,
  \dodoi{10.1126/science.1223344}

\bibitem[{Schneider {et~al.}(2015)Schneider, Izzard, Langer, \& {de
  Mink}}]{Schneider_etal.2015_EvolutionMassFunctions}
Schneider, F. R.~N., Izzard, R.~G., Langer, N., \& {de Mink}, S.~E. 2015, The
  Astrophysical Journal, 805, 20, \dodoi{10.1088/0004-637X/805/1/20}

\bibitem[{Sepinsky {et~al.}(2007)Sepinsky, Willems, Kalogera, \&
  Rasio}]{Sepinsky_etal.2007_InteractingBinariesEccentric}
Sepinsky, J.~F., Willems, B., Kalogera, V., \& Rasio, F.~A. 2007, The
  Astrophysical Journal, 667, 1170, \dodoi{10.1086/520911}

\bibitem[{Sepinsky {et~al.}(2009)Sepinsky, Willems, Kalogera, \&
  Rasio}]{Sepinsky_etal.2009_InteractingBinariesEccentric}
---. 2009, The Astrophysical Journal, 702, 1387,
  \dodoi{10.1088/0004-637X/702/2/1387}

\bibitem[{Sepinsky {et~al.}(2010)Sepinsky, Willems, Kalogera, \&
  Rasio}]{Sepinsky_etal.2010_InteractingBinariesEccentric}
---. 2010, Astrophysical Journal, 724, 546, \dodoi{10.1088/0004-637X/724/1/546}

\bibitem[{Soberman {et~al.}(1997)Soberman, Phinney, \& van~den
  Heuvel}]{Soberman_etal.1997_StabilityCriteriaMass}
Soberman, G.~E., Phinney, E.~S., \& van~den Heuvel, E. P.~J. 1997,
  arXiv:astro-ph/9703016.
\newblock \doeprint{astro-ph/9703016}

\bibitem[{Temmink {et~al.}(2022)Temmink, Pols, Justham, Istrate, \&
  Toonen}]{Temmink_etal.2022_CopingLossStability}
Temmink, K.~D., Pols, O.~R., Justham, S., Istrate, A.~G., \& Toonen, S. 2022,
  Coping with Loss: {{Stability}} of Mass Transfer from Post-Main Sequence
  Donor Stars

\bibitem[{{van Rossum}(1995)}]{vanRossum.1995_PythonTutorial}
{van Rossum}, G. 1995, Python Tutorial, Centrum voor Wiskunde en Informatica
  (CWI)

\bibitem[{{van Son} {et~al.}(2022){van Son}, {de Mink}, Renzo, Justham,
  Zapartas, Breivik, Callister, Farr, \&
  Conroy}]{vanSon_etal.2022_NoPeaksValleys}
{van Son}, L. A.~C., {de Mink}, S.~E., Renzo, M., {et~al.} 2022, No Peaks
  without Valleys: {{The}} Stable Mass Transfer Channel for Gravitational-Wave
  Sources in Light of the Neutron Star-Black Hole Mass Gap,  {arXiv},
  \dodoi{10.48550/arXiv.2209.13609}

\bibitem[{Woods \& Ivanova(2011)}]{Woods_Ivanova.2011_CanWeTrust}
Woods, T.~E., \& Ivanova, N. 2011, The Astrophysical Journal, 739, L48,
  \dodoi{10.1088/2041-8205/739/2/L48}

\end{thebibliography}

\appendix

\twocolumngrid

\section{correlated initial conditions}
\label{sec:impact_correlated_ICs}

To test the robustness of our results against variations in the initial conditions, we have implemented an alternative set of initial binary distributions based on the results of \citet{Moe_DiStefano.2017_MindYourPs}, who found that the orbital parameters of observed binaries are better reproduced by correlated distributions in the primary mass $M_1$, the mass ratio $q$, and the period $P$. They additionally identify an overall bias towards systems with short periods and unequal mass ratios, compared with the traditional distributions (used here) which are uniform in $q$ and loguniform in the orbital separation $a$  \citep{Sana_etal.2012_BinaryInteractionDominates}. 

Of course, these distributions are modelled after systems as they are observed today, which may deviate substantially from those at birth, but such corrections add further complications which are outside the scope of this paper. Furthermore, \citet{Moe_DiStefano.2017_MindYourPs} find that the average stellar multiplicity rises with increasing primary mass, such that primaries with $M\gtrsim~10~M_\odot$ are more likely than not to have two or more companions (see their Fig.~39). A similar conclusion about the correlation of multiplicity with either $P$ or $q$ cannot yet be drawn, due to the current lack of sufficient measurements of the masses and separations of the higher order bodies in hierarchical systems (Rosanne Di Stefano, private comm.). Multiplicity beyond binarity is not currently included in COMPAS.

\begin{figure}[t]
\centering
\includegraphics[width=0.97\columnwidth]{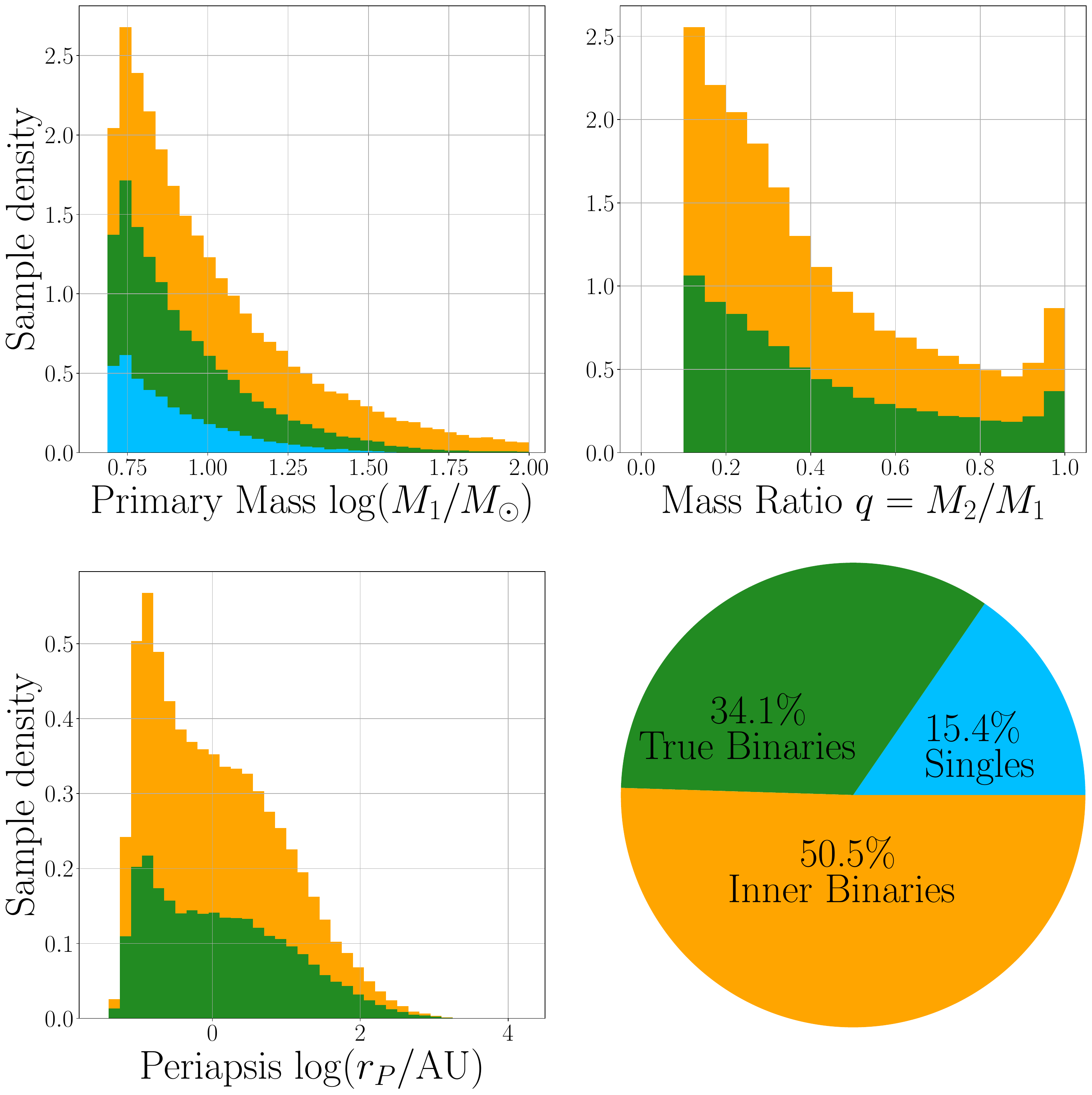}
\caption{
Initial condition distributions from \citet{Moe_DiStefano.2017_MindYourPs} for single stars (blue), isolated binaries (green), and binaries which are the inner components of hierarchical systems (orange).
Inner binaries make up more than half of all systems and are preferentially found at very tight separations with unequal mass ratios. 
} 
\label{fig:moeDiStef_ICs}
\end{figure}

Using the multiplicity correlation with primary mass, we can estimate the likelihood that a given sampled binary is in fact a \emph{true} binary, or is the inner binary of a higher multiplicity system. In Fig.~\ref{fig:moeDiStef_ICs}, we show the sampling distributions for the initial parameters, split into single stars, true binaries, and inner binaries. Inner binaries constitute a majority of the systems and dominate over true binaries at low mass ratios and small separations.

\begin{figure*}[!tb]
\centering
\includegraphics[width=.76\paperwidth]{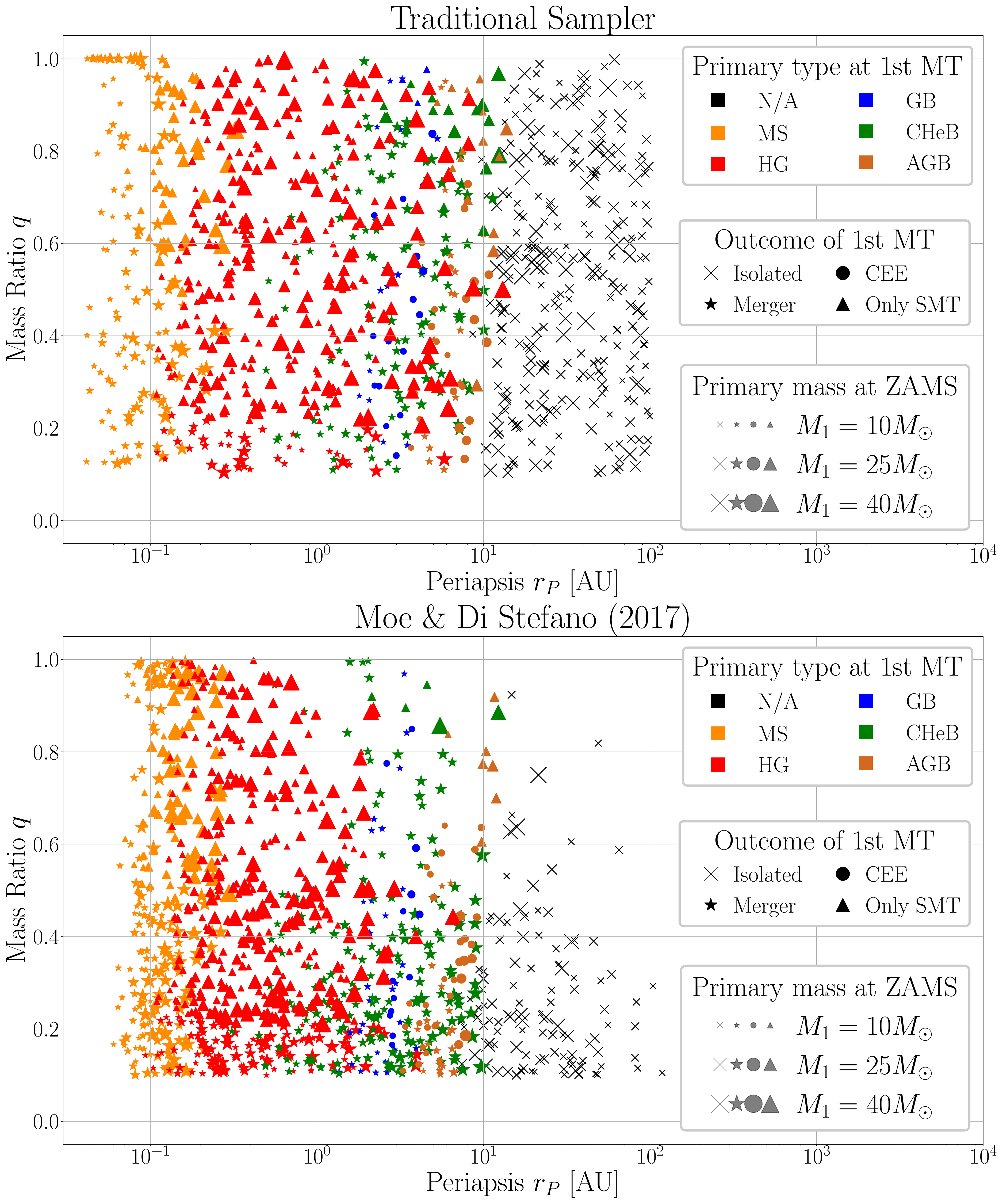}
\caption{
\textbf{The initial conditions and outcomes of the first mass transfer event} for binaries sampled from the Traditional Sampler and \citet{Moe_DiStefano.2017_MindYourPs} distributions. The plots show the sampled values for initial mass ratio and periapsis, with the size of the symbol indicating the initial primary mass. Periapsis is more indicative of future evolution than semi-major axis for the \citet{Moe_DiStefano.2017_MindYourPs} sampler (which includes non-zero eccentricities), because we circularize about the periapsis when \ac{MT} starts (see text). The marker shape indicates the outcome of the first binary interaction. The color represents the stellar type of the primary when it first interacts with its companion. Non-interacting binaries are indicated with a black X. The cluster of short period, equal mass ratio binaries from the top plot corresponds to over-contact binaries, which are then equilibrated to have equal masses.
} 
\label{fig:uniform_vs_moedistef_PQs}
\end{figure*}

Samples drawn from the Traditional Sampler and \citet{Moe_DiStefano.2017_MindYourPs} distributions are presented for comparison in Fig.~\ref{fig:uniform_vs_moedistef_PQs}. Each point in these plots represents a simulated binary, with initial orbital periapsis $r_P$ and mass ratio $q$ indicated by the abscissa and ordinate, respectively.  The initial primary mass is represented by the size of the point. The colors correspond to the primary stellar type immediately prior to the first \ac{MT} interaction (if any), with orange, red, blue, green, and pink corresponding to Main Sequence (MS), Hertzsprung Gap (HG), Giant Branch (GB), Core Helium Burning (CHeB), and Asymptotic Giant Branch (AGB) stars, respectively. The shapes indicate the outcome of this interaction: a star for stellar mergers, a circle for \ac{CEE} survivors, and a triangle for stable \ac{MT}. Non-interacting binaries are represented by a black X. Both plots use the default evolutionary model. 

We emphasize that these two initial condition distributions are not meant to be representative of the large parameter space of possible input distributions. Both distributions, for example, assume a Kroupa \ac{IMF} for the primary mass, though the universality of the \ac{IMF} is a matter of ongoing debate \citep{Grudic_etal.2023_DoesGodPlay}. Nevertheless, their substantial differences provide a useful comparison. 

Surprisingly, the rates of different mass transfer outcomes, including the stable and unstable \ac{MT} channels and stellar mergers, showed a surprising \emph{insensitivity} to differences in the initial distributions. This insensitivity holds even across variations to the \ac{MT} efficiency, \ac{AM} loss, and the donor stability criteria. However, a more careful treatment of higher multiplicity systems may result in a greater sensitivity to the initial conditions. 

\end{document}